\documentclass[twocolumn,aps,prd,superscriptaddress,nofootinbib,floatfix]{revtex4-1}

\usepackage{graphicx}
\usepackage{multirow}
\usepackage{hyperref}
\usepackage{breakurl}

\newcommand{\nn}{\nonumber}
\newcommand{\beq}{\begin{equation}}
\newcommand{\eeq}{\end{equation}}
\newcommand{\beqa}{\begin{eqnarray}}
\newcommand{\eeqa}{\end{eqnarray}}
\newcommand{\TeV}{\rm TeV}

\def\rhobar{\bar\rho}
\def\etabar{\bar\eta}

\newcommand{\Bbar}{\,\overline{\!B}{}}
\newcommand{\Dbar}{\,\overline{\!D}{}}
\newcommand{\Kbar}{\,\overline{\!K}{}}
\def\B0bar{\Bbar{}^0}
\def\D0bar{\Dbar{}^0}
\def\K0bar{\Kbar{}^0}

\begin{document}

\title{\boldmath Future sensitivity to new physics in $B_d$, $B_s$ and $K$ mixings}

\author{J\'er\^ome Charles$^*$}
\affiliation{Aix-Marseille Universit\'e, CNRS, CPT, UMR 7332, 13288 Marseille, France}
\affiliation{Universit\'e de Toulon, CNRS, CPT, UMR 7332, 83957 La Garde, France}

\author{S\'ebastien Descotes-Genon$^*$}
\affiliation{Laboratoire de Physique Th\'eorique, CNRS/Univ.\ Paris-Sud 11 
(UMR 8627) 91405 Orsay Cedex, France}

\author{Zoltan Ligeti}
\affiliation{Ernest Orlando Lawrence Berkeley National Laboratory,
University of California, Berkeley, CA 94720}

\author{St\'ephane Monteil$^*$}
\affiliation{Laboratoire de Physique Corpusculaire de Clermont-Ferrand
Universit\'e Blaise Pascal\\
24 Avenue des Landais F-63177 Aubiere Cedex}

\author{Michele Papucci}
\affiliation{Michigan Center for Theoretical Physics, 
University of Michigan, Ann Arbor, MI 48109}

\author{Karim Trabelsi$^*$}
\affiliation{\hbox{High Energy Accelerator Research Organization, KEK
1-1 Oho, Tsukuba, Ibaraki 305-0801, Japan}\\
$^*$for the CKMfitter Group}

\begin{abstract}

We estimate, in a large class of scenarios, the sensitivity to new physics in
$B_d$ and $B_s$ mixings achievable with 50\,ab$^{-1}$ of Belle~II and
50\,fb$^{-1}$ of LHCb data.  We find that current limits on new physics
contributions in both $B_{d,s}$ systems can be improved by a factor of $\sim$\,5
for all values of the $CP$ violating phases, corresponding to over a factor of 2
increase in the scale of new physics probed.  Assuming the same suppressions by
CKM matrix elements as those of the standard model box diagrams, the scale
probed will be about 20\,TeV for tree-level new physics contributions, and about
2\,TeV for new physics arising at one loop.  We also explore the future
sensitivity to new physics in $K$ mixing.  Implications for generic new physics
and for various specific scenarios, such as minimal flavor violation, light
third-generation dominated flavor violation, or $U(2)$ flavor models are
studied.

\end{abstract}

\maketitle

\section{Introduction}\label{sec:intro}

Before the impressive results from the $B$ factory experiments, BaBar and Belle,
the simple picture of Kobayashi and Maskawa for the origin of the $CP$
violation~\cite{Kobayashi:1973fv} observed in $K$ decays was not confirmed
experimentally.  The BaBar and Belle results showed that the SM description of
the flavor sector is correct at the order one level.  However, in most
flavor-changing neutral-current processes, new physics (NP) can still contribute
at least at the level of 20--30\% compared to the SM.

Many extensions of the SM receive stringent constraints from data on flavor
changing processes and $CP$ violation, and may give observable effects as the
sensitivity improves.  The mixings of the four neutral mesons, $K$, $D$, $B_d$,
and $B_s$, provide particularly strong bounds.  For each neutral-meson system,
contributions generated by new heavy degrees of freedom can be described by
two real parameters.  For example, in low-energy supersymmetry $B$ mixing
receives contributions (besides the SM box diagrams with $W$ bosons and top
quarks) from box diagrams with winos and stops or gluinos and sbottoms.  The
magnitudes and phases of such contributions depend crucially on the mechanism of
supersymmetry breaking and the origin of flavor symmetry breaking.  

However, the extraction of NP contribution to meson mixing is entangled with the
determination of the SM parameters, in particular the CKM elements. It is not
enough to measure the mixing amplitude itself, only the combination of many
measurements can reveal a deviation from the SM.  In this paper we perform such
a fit, taking into account the latest expectations for future LHCb and Belle~II
measurements, and anticipated progress in lattice QCD, in order to investigate
the sensitivity to NP in neutral-meson mixing in the near future.

In most of this paper, we consider the well-defined scenario where no deviations
from the SM predictions are observed. This allows us to explore the expected
progress in constraining NP in the mixings of neutral mesons in an unambiguous
way.   An illustration of the prospects to reveal a possible NP signal is given
in the last section. 

\begin{table*}
{\scriptsize
\begin{tabular}{c|cccccc}
\hline\hline
  &  2003  &  2013  &  Stage~I  &  & Stage~II & \\
\hline\hline
$|V_{ud}|$  &  $0.9738\pm 0.0004$  &  $0.97425\pm 0\pm 0.00022$  &  id  &  & id  &  \\
$|V_{us}|$ $(K_{\ell 3})$  &  $0.2228\pm 0.0039\pm 0.0018$  &  $0.2258\pm 0.0008
\pm 0.0012$  &  
  $0.22494\pm 0.0006$  &  & id  &  \\
$|\epsilon_K|$  &  $(2.282 \pm 0.017) \times 10^{-3}$  &  $(2.228 \pm 0.011)\times 10^{-3}$  &  id  &  &
 id  &  \\
$\Delta m_d$ [${\rm ps}^{-1}$]  &  $0.502\pm 0.006$  &  $0.507\pm 0.004$  &  id  &  & id &\\
$\Delta m_s$ [${\rm ps}^{-1}$]  &  $>14.5$ [95\% CL]   &  $17.768\pm 0.024$  &  id   &  & id &\\
$|V_{cb}|\times 10^3$  ($b\to c\ell\bar\nu$) &  $41.6 \pm 0.58 \pm 0.8$  &  $41.15 \pm 0.33 \pm 0.59$  &   
 $42.3\pm 0.4$& \cite{Aushev:2010bq} &
  $42.3\pm 0.3$& \cite{Aushev:2010bq}\\
$|V_{ub}|\times 10^3$ ($b\to u\ell\bar\nu$)  &  $3.90 \pm 0.08 \pm 0.68$  &  $3.75 \pm 0.14 \pm 0.26$  &  
  $3.56\pm 0.10$& \cite{Aushev:2010bq} &
  $3.56\pm 0.08$& \cite{Aushev:2010bq} \\
\hline
$\sin 2\beta$  &  $0.726\pm 0.037$  &  $0.679\pm 0.020$  &  $0.679\pm 0.016$
  &  \cite{Aushev:2010bq}  &  $0.679\pm 0.008$  &  \cite{Aushev:2010bq}\\
$\alpha$ (mod $\pi$) &  ---  &  $(85.4^{+4.0}_{-3.8})^\circ$  &  $(91.5\pm 2)^\circ$  &  \cite{Aushev:2010bq}  &  $(91.5\pm 1)^\circ$& \cite{Aushev:2010bq}\\
$\gamma$ (mod $\pi$)  &  ---  &  $(68.0^{+8.0}_{-8.5})^\circ$  &  $(67.1\pm 4)^\circ$  &  \cite{Bediaga:2012py, Aushev:2010bq}  
  &  $(67.1\pm 1)^\circ$  &  \cite{Bediaga:2012py, Aushev:2010bq}\\
$\beta_s$  &  ---  &  $0.0065^{+0.0450}_{-0.0415}$  &  
  $0.0178\pm 0.012$  &  \cite{Bediaga:2012py}& $0.0178\pm 0.004$  &  \cite{Bediaga:2012py} \\
${\cal B}(B\to\tau\nu)\times 10^4$  &  ---  &  $1.15\pm 0.23$
  &  $0.83\pm 0.10$  &  \cite{Aushev:2010bq}
  &  $0.83\pm 0.05$  &  \cite{Aushev:2010bq}\\
${\cal B}(B\to\mu\nu)\times10^7$  &  ---  &  ---  
  &  $3.7 \pm 0.9$   &  \cite{Aushev:2010bq} & $3.7\pm 0.2$  &  \cite{Aushev:2010bq}\\
$A_{\rm SL}^d \times 10^4$  &  $10 \pm 140$  &  $ 23\pm 26$  &  $-7\pm 15$ & \cite{Aushev:2010bq} & $-7\pm 10$ & \cite{Aushev:2010bq}\\
$A_{\rm SL}^s \times 10^4$  &  ---  &  $ -22\pm 52$  &  $0.3\pm 6.0$ & \cite{Bediaga:2012py} & $0.3\pm 2.0$ & \cite{Bediaga:2012py}\\
\hline
$\bar{m}_c$  &  $1.2\pm 0 \pm 0.2$  &  $1.286\pm 0.013 \pm 0.040$  &  $1.286\pm 0.020$ &  & $1.286\pm 0.010$ &\\
$\bar{m}_t$  &  $167.0\pm 5.0$  &  $165.8\pm 0.54\pm0.72$  &  id  &  & id &\\
$\alpha_s(m_Z)$  &  $0.1172\pm 0 \pm 0.0020$  &  $0.1184\pm 0\pm 0.0007$   &  id && id &\\
$B_K$  &  $0.86\pm 0.06\pm 0.14$  &  $0.7615\pm 0.0026\pm 0.0137$  &  $0.774\pm 0.007$  &  \cite{USlqcd,Ruth} & $0.774\pm 0.004$ & \cite{USlqcd,Ruth}\\
$f_{B_s}$ [GeV]  &  $0.217\pm 0.012\pm 0.011$  &  $0.2256\pm 0.0012\pm 0.0054$  &  $0.232\pm 0.002$  & \cite{USlqcd,Ruth} & $0.232\pm 0.001$  & \cite{USlqcd,Ruth}\\
$B_{B_s}$  &  $1.37\pm 0.14$  &  $1.326\pm 0.016\pm 0.040$  &  $1.214 \pm 0.060$  & \cite{USlqcd,Ruth} & $1.214 \pm 0.010$   & \cite{USlqcd,Ruth} \\
$f_{B_s}/f_{B_d}$  &  $1.21\pm 0.05\pm 0.01$  &  $1.198\pm 0.008\pm 0.025$  &  $1.205\pm 0.010$  & \cite{USlqcd,Ruth}  & $1.205\pm 0.005$ & \cite{USlqcd,Ruth}\\
$B_{B_s}/B_{B_d}$  &  $1.00\pm 0.02$  &  $1.036\pm 0.013\pm 0.023$  &  $1.055\pm 0.010$  & \cite{USlqcd,Ruth} &   $1.055\pm 0.005$ & \cite{USlqcd,Ruth}\\
$\tilde{B}_{B_s}/\tilde{B}_{B_d}$  &  ---  &  $1.01\pm 0 \pm 0.03$  &   $1.03 \pm 0.02$  &  & id  &   \\
$\tilde{B}_{B_s}$  &  ---  &  $0.91\pm 0.03\pm 0.12$  &  $0.87 \pm 0.06$  &  & id
& \\
\hline\hline
\end{tabular}}
\caption{Central values and uncertainties used in our analysis (see definitions
in Ref.~\cite{Lenz:2010gu}).  The entries ``id" refer to the value in the same
row in the previous column. The 2003 and 2013 values correspond to Lepton-Photon
2003 and FPCP 2013 conferences~\cite{Charles:2004jd}. The assumptions entering
the Stage~I and Stage~II estimates are described in the text.}
\label{bigtable}
\end{table*}

\section{new physics in meson mixing}
\label{sec:mixing}

In a large class of NP models the unitarity of the CKM matrix is maintained, and
the most significant NP effects occur in observables that vanish at tree level
in the SM.  In the SM CKM fit, the constraints come from (i) $\Delta F=1$
processes dominated by tree-level charged current interactions, and (ii) $\Delta
F=2$ meson mixing processes, which only arise at loop level.  Therefore, it is
simple to modify the CKM fit to constrain new physics in $\Delta F=2$ processes,
under the assumption that it does not significantly affect the SM tree-level
charged-current interactions~\cite{Soares:1992xi}.  Within this framework (for a
review, see~\cite{Hocker:2006xb}), we can parameterize the NP contributions to
the $B_{d,s}$ mixing amplitudes as
\beq\label{param}
M_{12}^{d,s} = (M_{12}^{d,s})_{\rm SM} \times
  \big(1 + h_{d,s}\, e^{2i\sigma_{d,s}}\big)\,.
\eeq
Until the first measurements of $\alpha$ and $\gamma$ around 2003, it was not
known if the SM gives the leading contribution to $B_d$\,--\,$\Bbar_d$
mixing~\cite{Charles:2004jd, Ligeti:2004ak} (similarly, for
$B_{s}$\,--\,$\Bbar_{s}$ mixing, the LHCb constraint on $\sin 2\beta_{s}$ was
needed).

The motivation for the above parameterization is that any NP
contribution to $M_{12}$ is additive, and using Eq.~(\ref{param}) one can easily
read off both the magnitude and the $CP$ violating phase of the total NP
contribution.  In particular, for a NP contribution to the mixing of a
meson with $q_i\bar q_j$ flavor quantum numbers due to the operator
\beq\label{operator}
\frac{C_{ij}^2}{\Lambda^2}\, (\bar q_{i,L}\gamma^{\mu}q_{j,L})^2\,,
\eeq
one finds that 
\begin{eqnarray}\label{hnumeric}
h &\simeq & 1.5\, \frac{|C_{ij}|^2}{|\lambda^{t}_{ij}|^2}\,
  \frac{(4\pi)^2}{G_F\Lambda^2} \simeq \frac{|C_{ij}|^2}{|\lambda^t_{ij}|^{2}} 
  \left(\frac{4.5\, \TeV}{\Lambda}\right)^2 , \nn\\ 
\sigma &=& {\rm arg}\big(C_{ij}\, \lambda_{ij}^{t*}\big) ,
\end{eqnarray}
where $\lambda^{t}_{ij} = V_{ti}^{*}\, V_{tj}$ and $V$ is the CKM matrix. 
We used NLO expressions for the SM and LO for NP, and neglected running for NP
above the top mass.
Operators of different chiralities have conversion factors differing by ${\cal
O}(1)$ factors~\cite{Buras:2001ra}. Minimal flavor violation (MFV), where the
NP contributions are aligned with the SM ones, correspond to
$\sigma = 0$ (mod $\pi/2$).

Analogously, in $K$ mixing, we choose to parameterize NP via an additive term to
the so-called $tt$ contribution to $M_{12}^{K}$ in the SM. This is justified by
the short distance nature of NP, by the fact that in many NP models the largest
contribution to $M_{12}^{K}$ arise mostly via effects involving the third
generation (``23--31'' mixing), and more practically, since this allows one to
maintain a consistent normalization for NP across the three down-type neutral
meson systems. In this paper, $D$-meson mixing is not considered, due to the
large uncertainties related to long-distance contributions.

Comments are in order concerning our assumption of neglecting NP in charged
current $b\to u,c$ transitions.  If a NP contamination is present and has a
different chiral structure than the SM, it will manifest itself by modifying
decay distributions, such as the lepton spectrum in semileptonic $B$ decays.  On
the contrary, if NP has the same chiral structure as the SM, it cannot be
physically separated in the determination of $\rhobar$ and $\etabar$. In such a
case, the extracted values of these parameters will not correspond to their SM
values.  This discrepancy will propagate to the NP fit, and will manifest itself
as a nonzero value for $h_{d,s}$~\cite{Fox:2007in}, with a specific pattern for
$h_{d,s}$ and $\sigma_{d,s}$.

\section{\boldmath Generic fit for $B_d$ and $B_s$ mixings}
\label{sec:fit}

\begin{figure*}[tb]
\includegraphics[width=.45\textwidth,clip,bb=15 15 550 520]{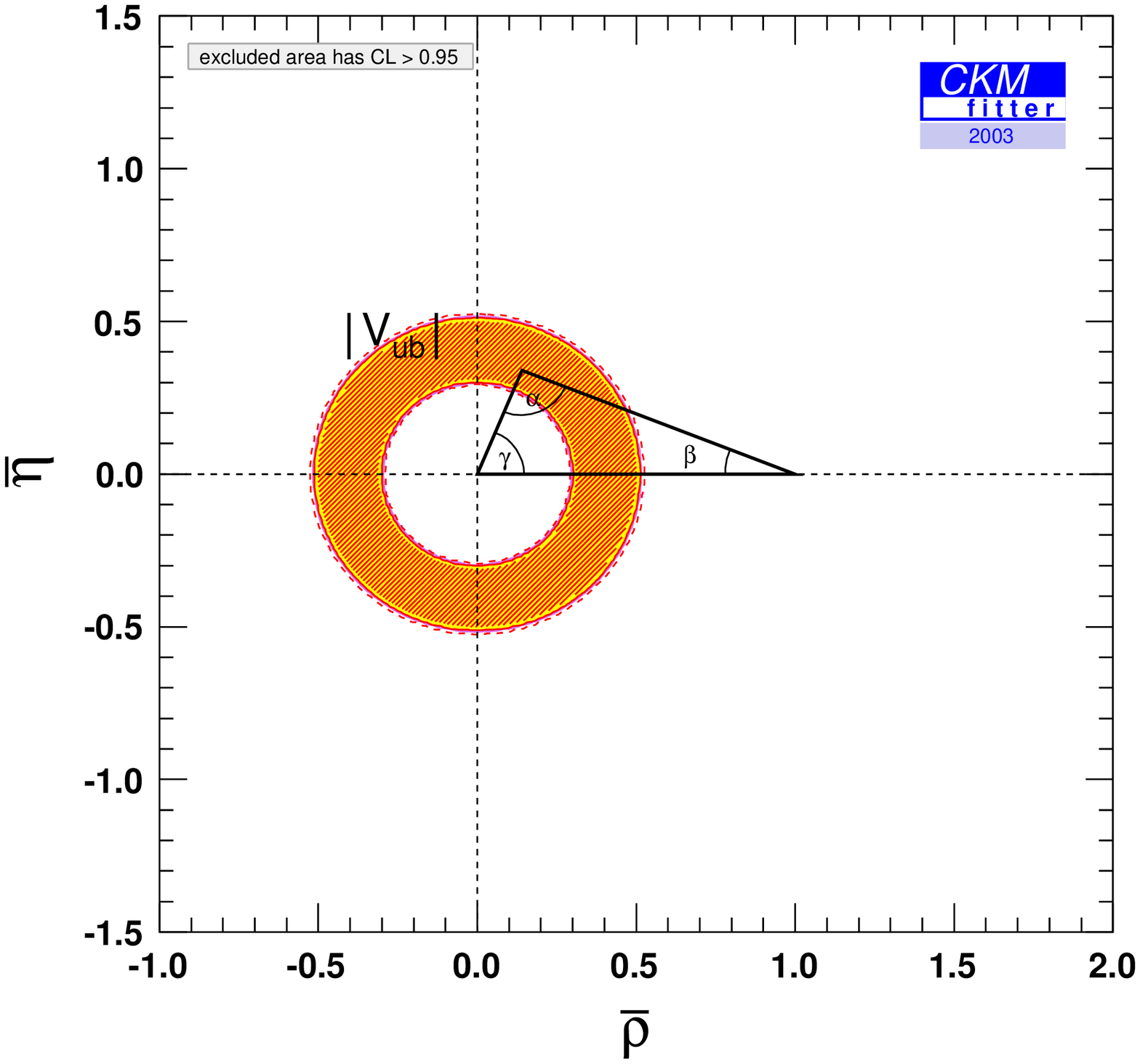}\hfil
\includegraphics[width=.45\textwidth,clip,bb=15 15 550 520]{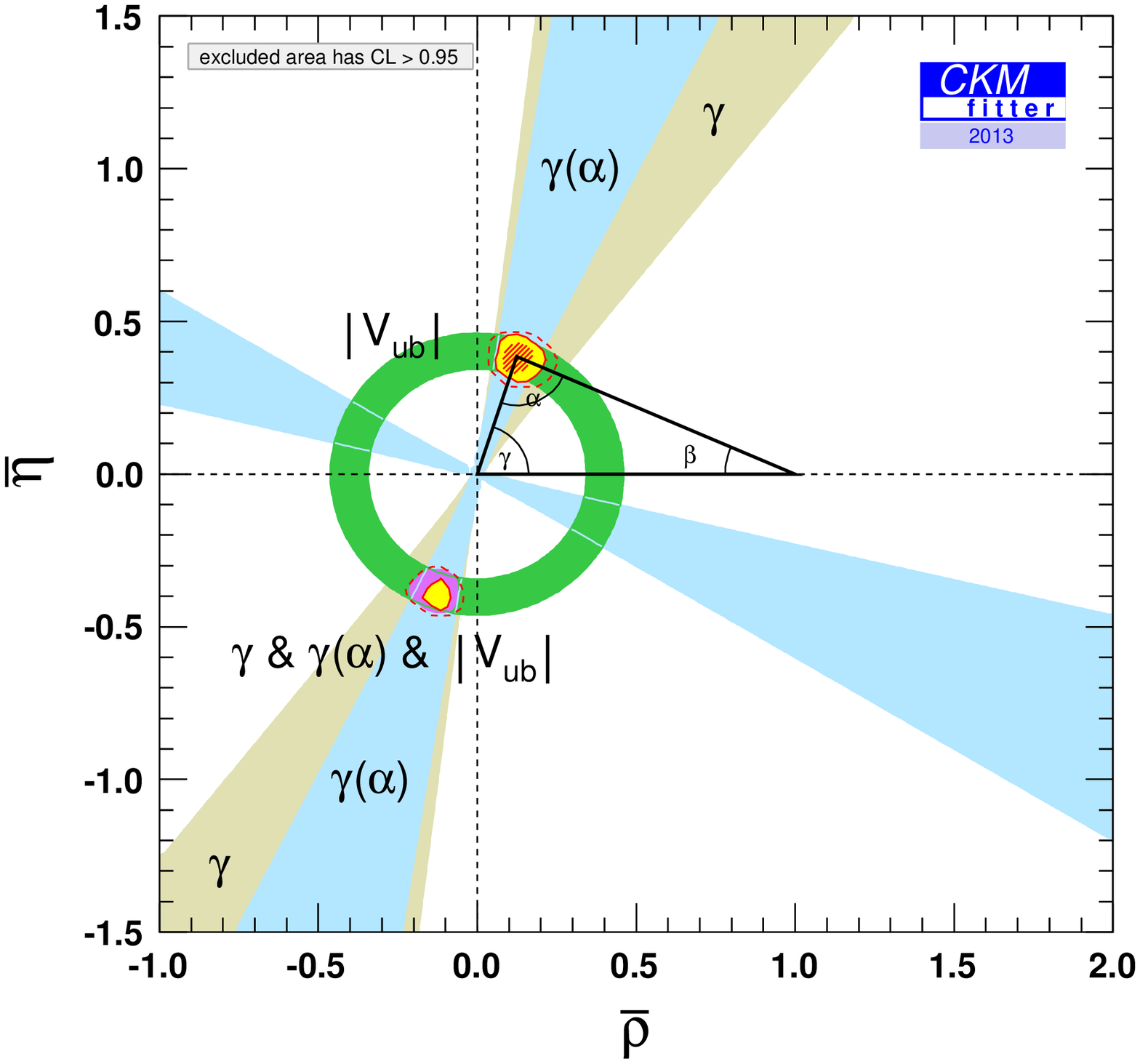}
\includegraphics[width=.45\textwidth,clip,bb=103 20 670 550]{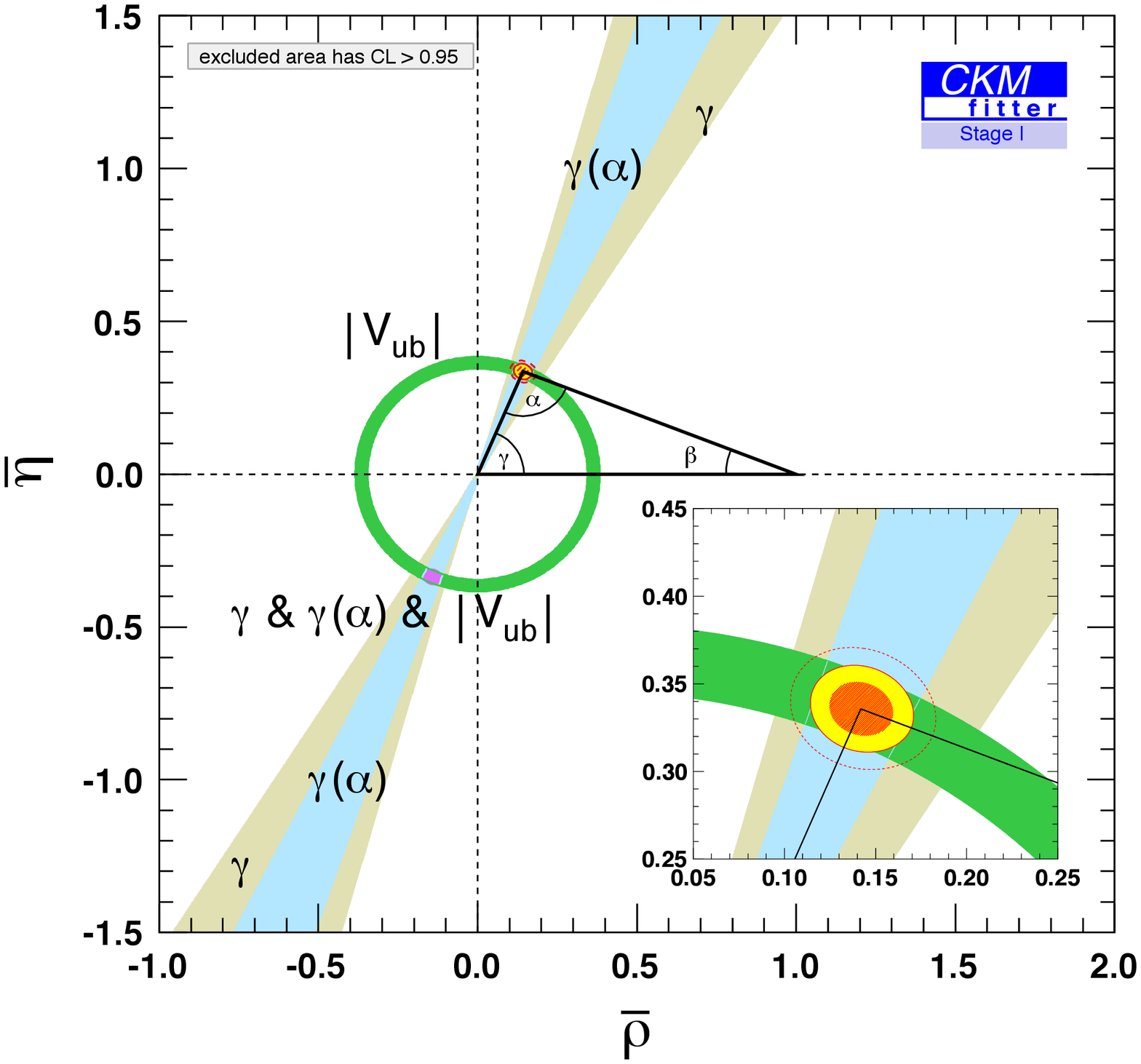}\hfil
\includegraphics[width=.45\textwidth,clip,bb=103 20 670 550]{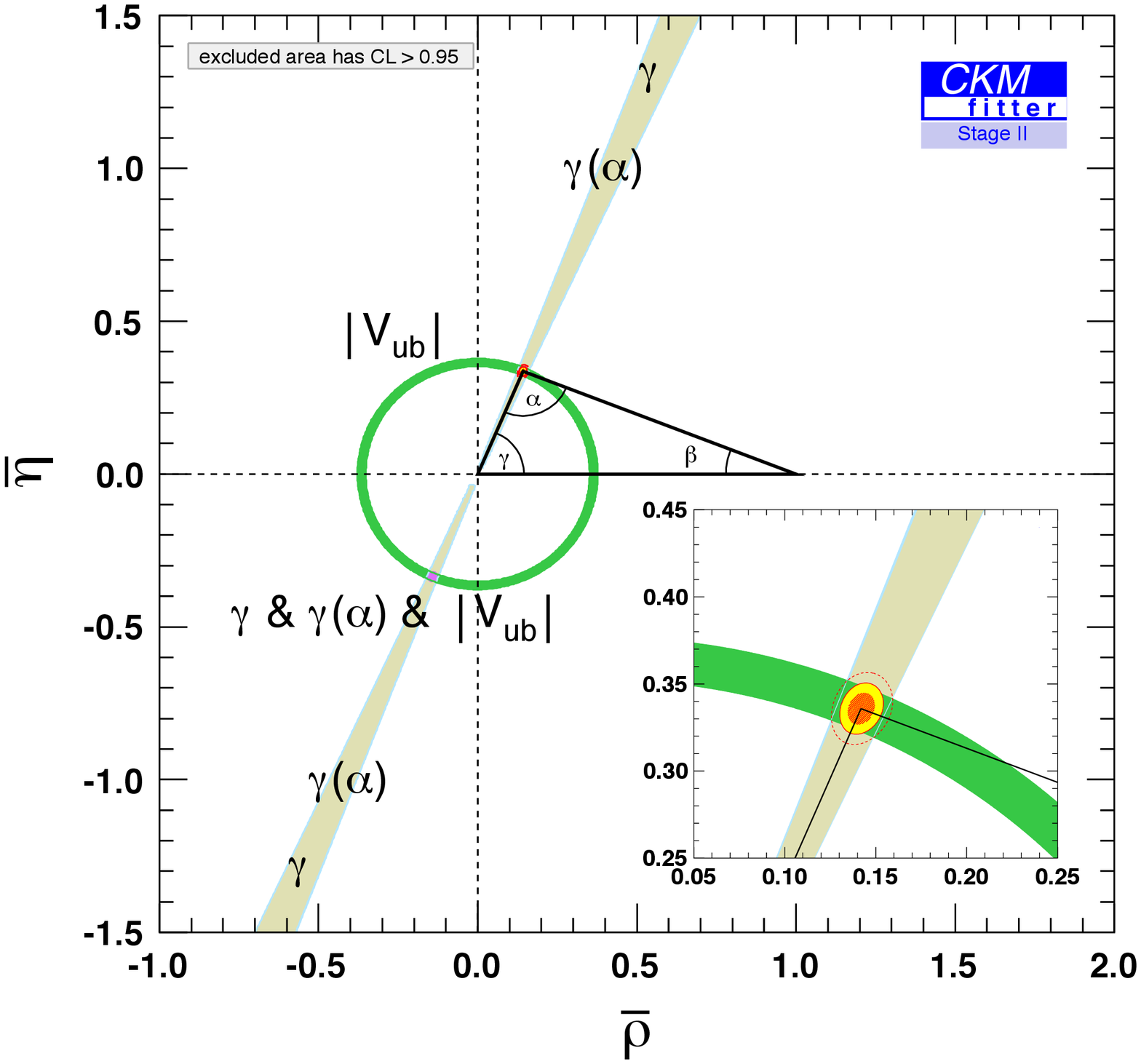}
\caption{The past (2003, top left) and present (top right) status of the
unitarity triangle in the presence of NP in neutral-meson mixing. The lower
plots show future sensitivities for Stage~I and Stage~II described in the text,
assuming data consistent with the SM. The combination of all constraints in
Table~\ref{bigtable} yields the red-hatched regions, yellow regions, and dashed
red contours at 68.3\%\,CL, 95.5\%\,CL, and 99.7\%\,CL, respectively.}
\label{CKMfit}
\end{figure*}

\begin{figure*}[tb]
\includegraphics[width=.48\textwidth,clip,bb=15 15 550 470]{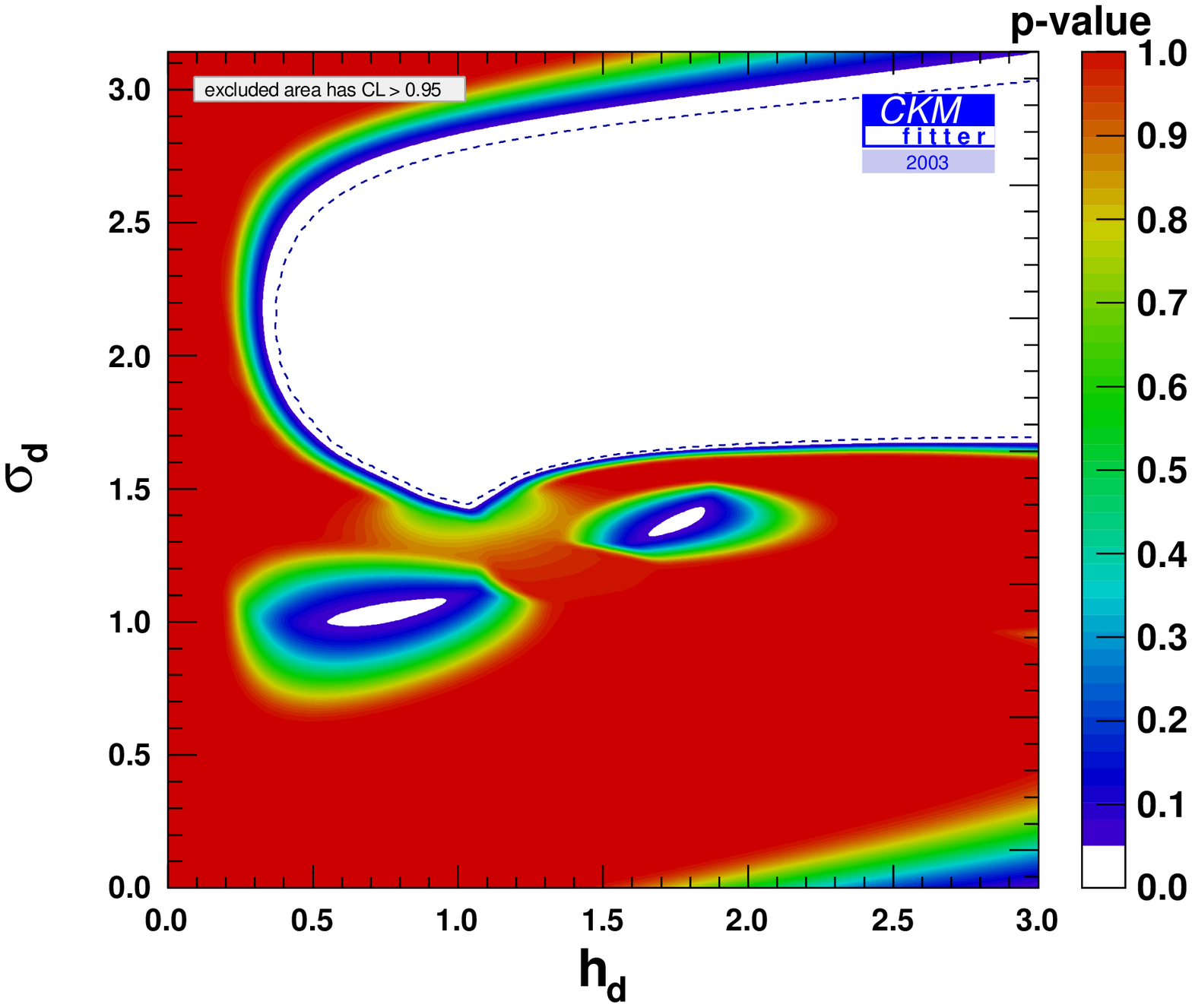}\hfill
\includegraphics[width=.48\textwidth,clip,bb=15 15 550 470]{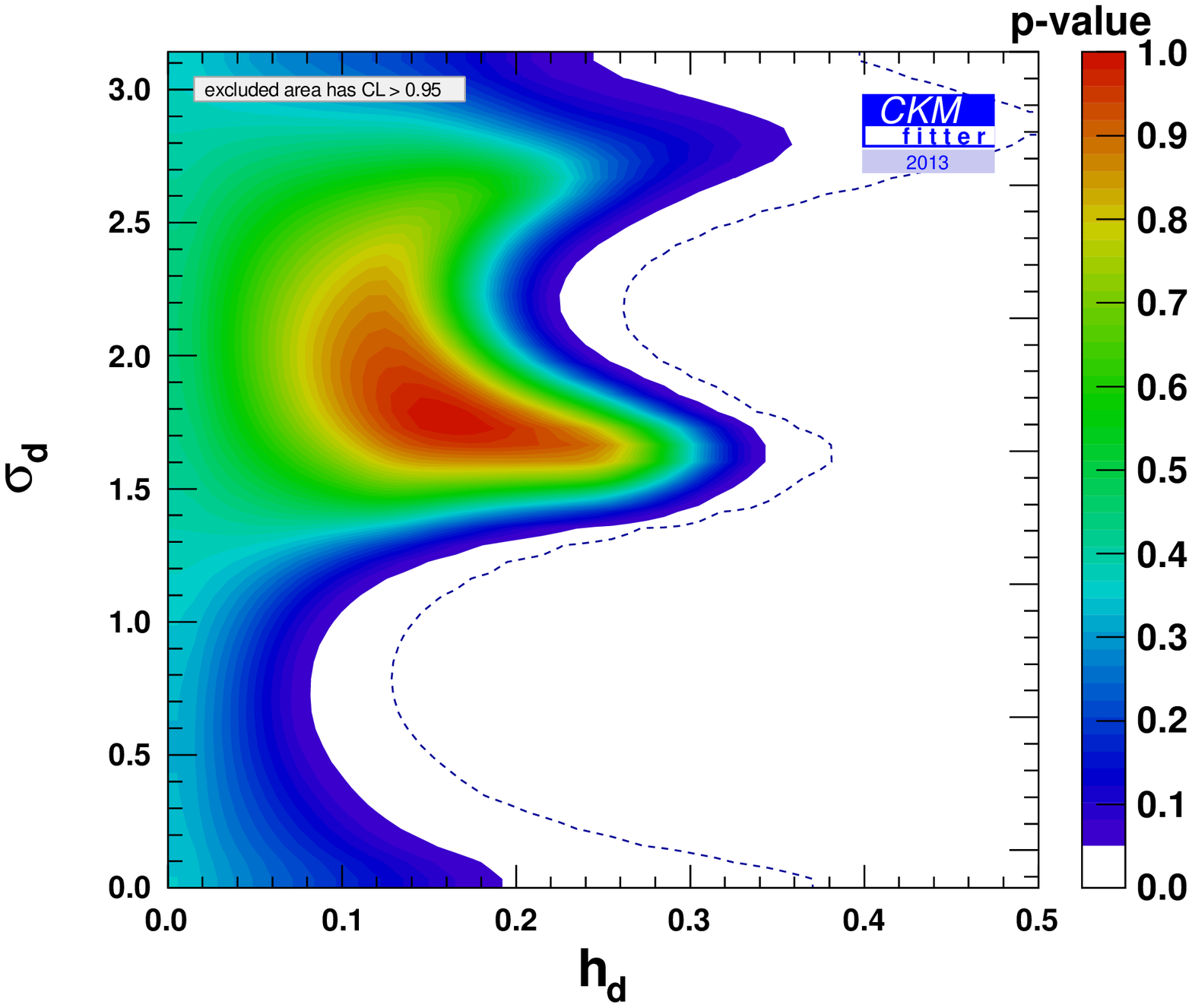}
\includegraphics[width=.48\textwidth,clip,bb=15 15 550 470]{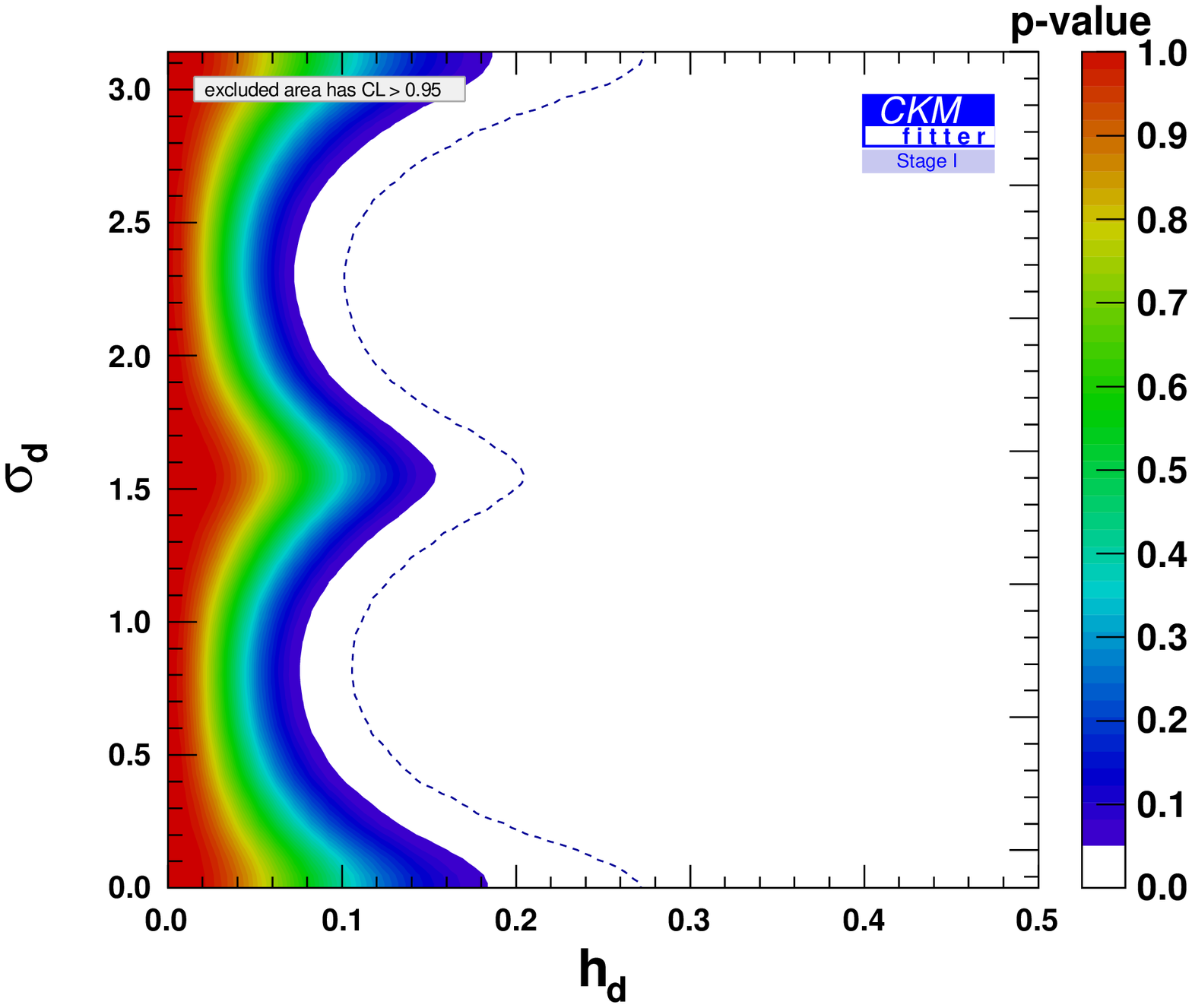}\hfill
\includegraphics[width=.48\textwidth,clip,bb=15 15 550 470]{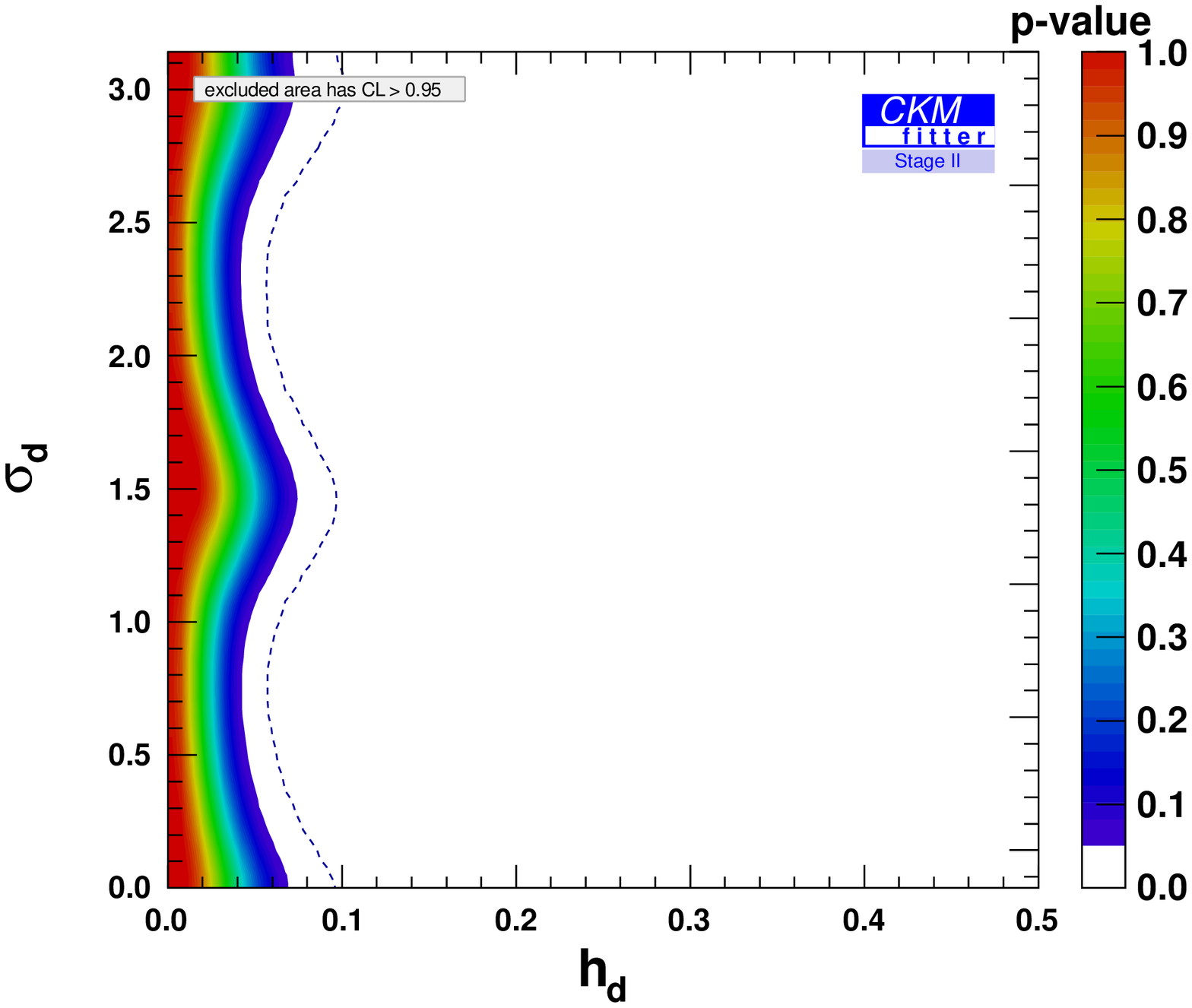}
\caption{The past (2003, top left) and present (top right) constraints on
$h_d-\sigma_d$ in $B_d$ mixing. The lower plots show future sensitivities for
Stage~I and Stage~II described in the text, assuming measurements consistent
with the SM. The dotted curves show the 99.7\%\,CL contours.}
\label{Bdfit}
\end{figure*}

\begin{figure*}[tb]
\includegraphics[width=.48\textwidth,clip,bb=15 15 550 470]{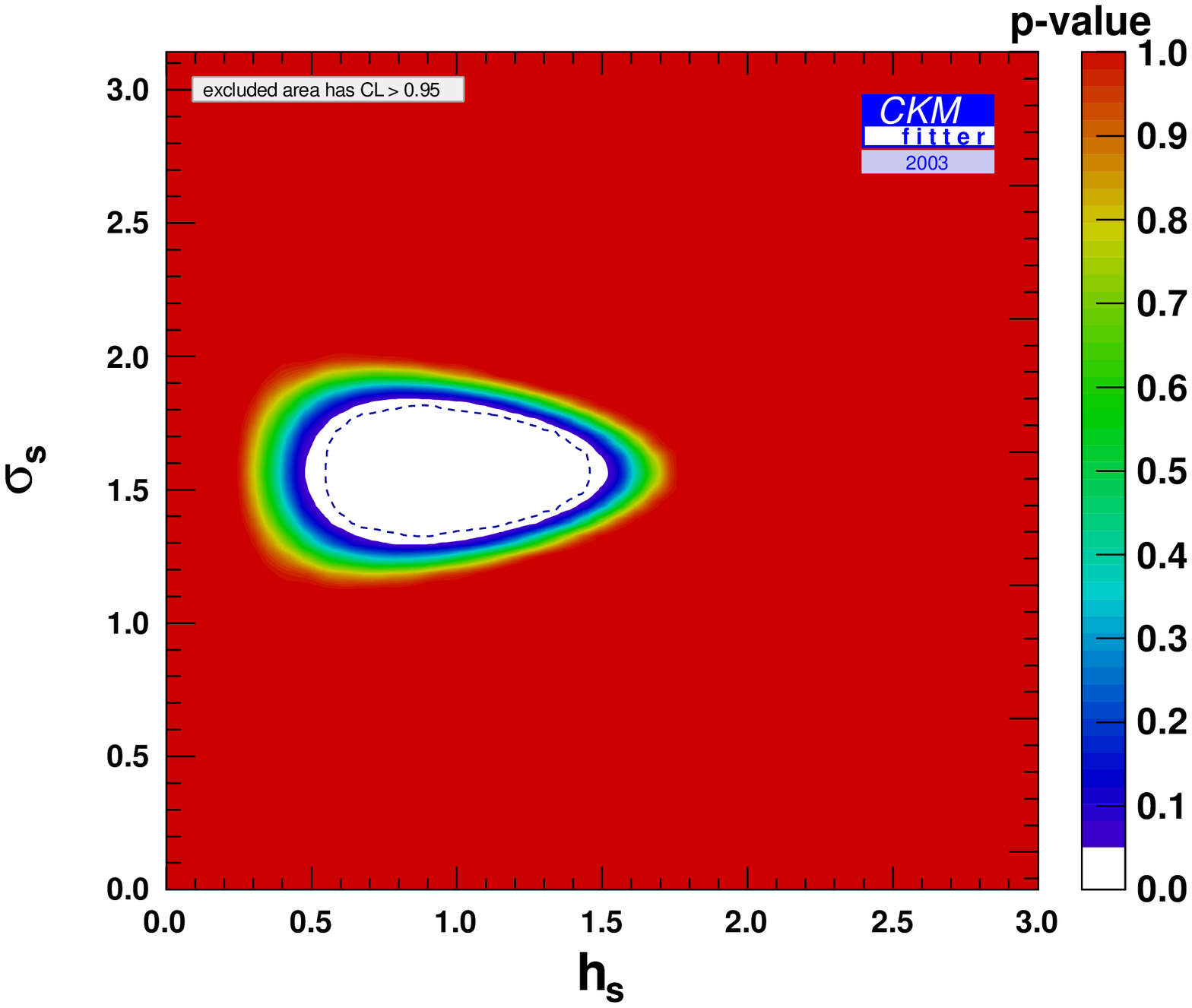}\hfill
\includegraphics[width=.48\textwidth,clip,bb=15 15 550 470]{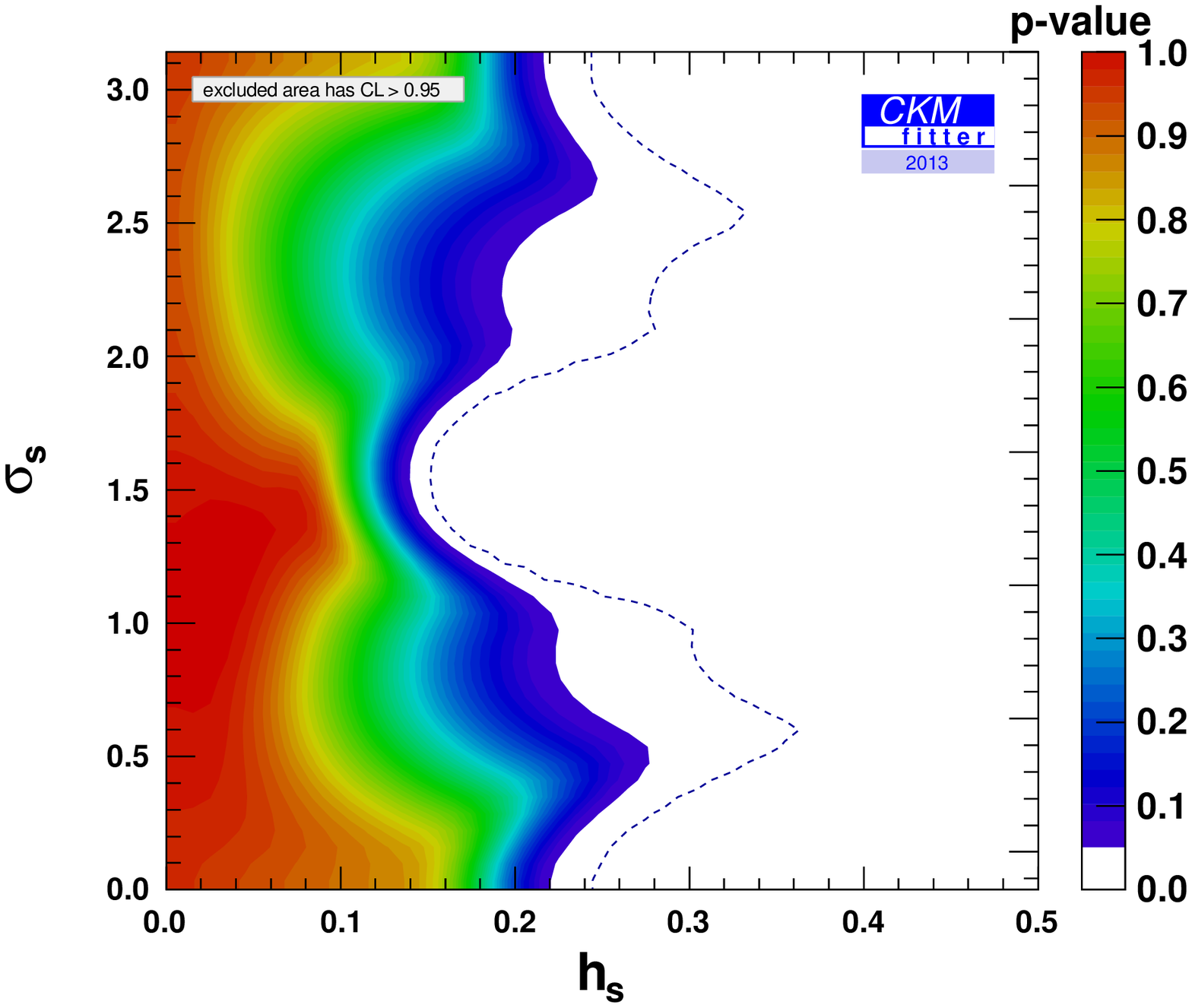}
\includegraphics[width=.48\textwidth,clip,bb=15 15 550 470]{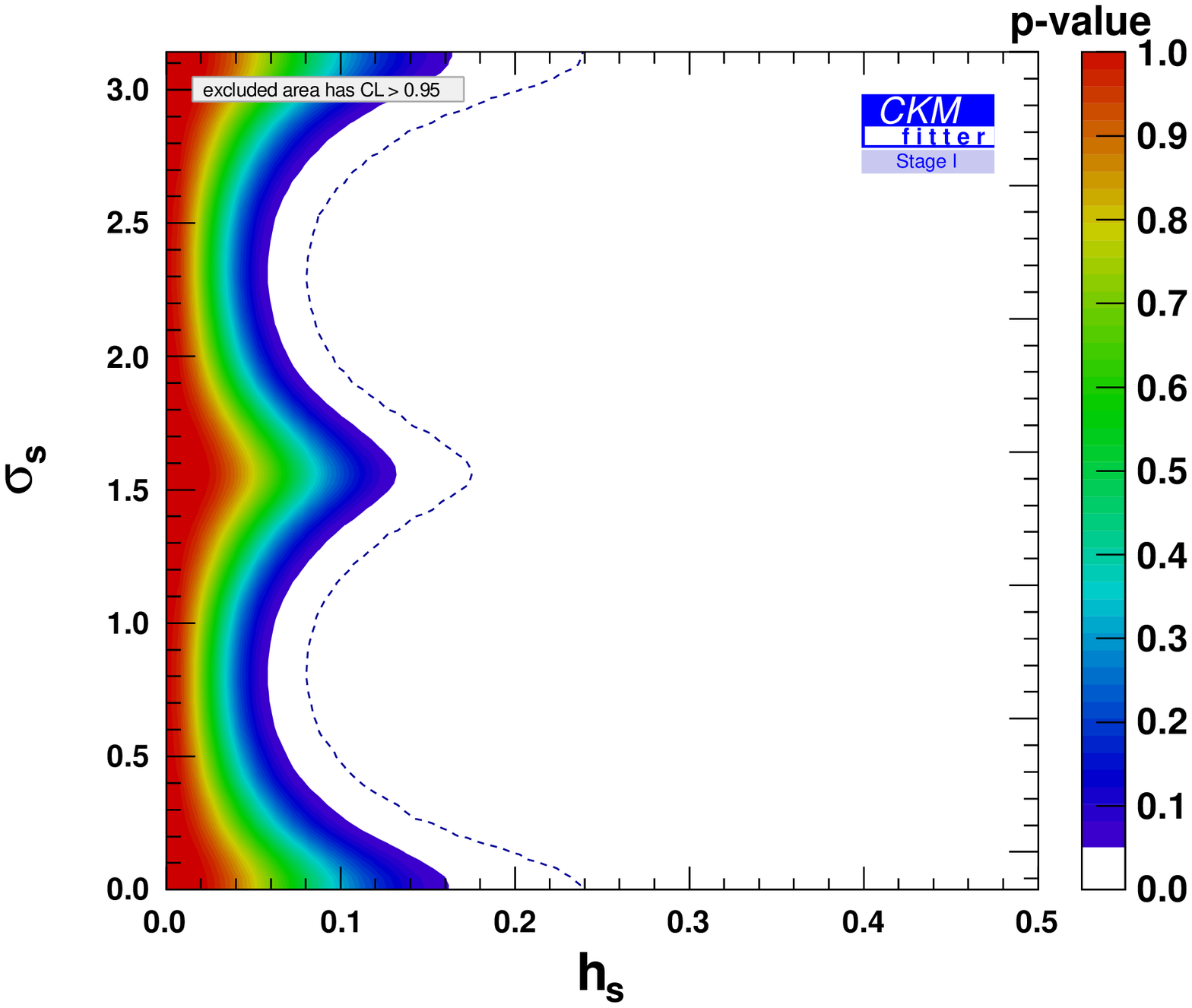}\hfill
\includegraphics[width=.48\textwidth,clip,bb=15 15 550 470]{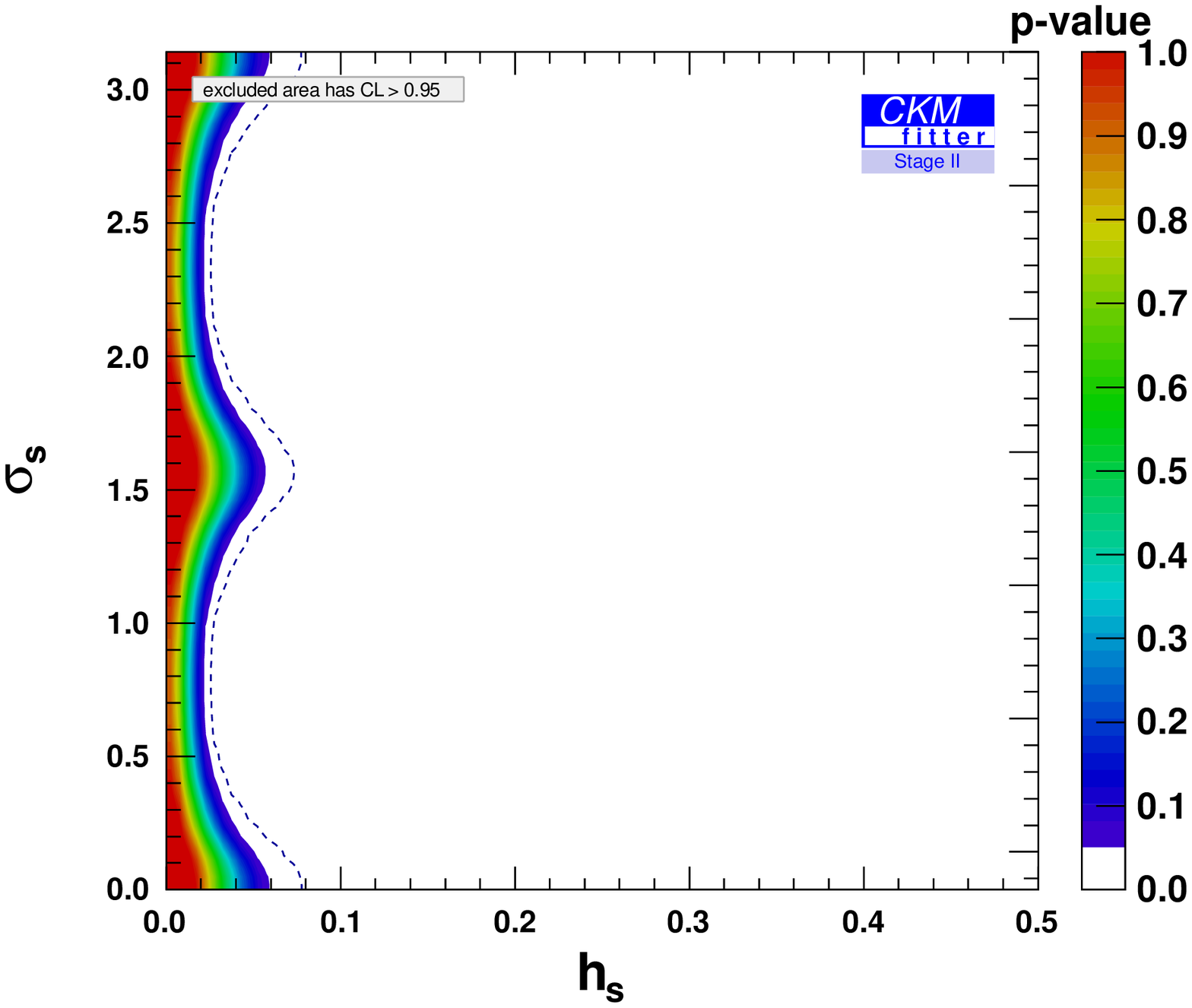}
\caption{The past (2003, top left) and present (top right) constraints on
$h_s-\sigma_s$ in $B_s$ mixing. The lower plots show future sensitivities for
the Stage~I and Stage~II described in the text, assuming measurements consistent
with the SM. The dotted curves show the 99.7\%\,CL contours.}
\label{Bsfit}
\end{figure*}

Table~\ref{bigtable} shows all inputs and their uncertainties used in our fit,
performed using the CKMfitter package~\cite{ckmfitter, Charles:2004jd,
Charles:2011va} with its extension to NP in $\Delta F=2$~\cite{Lenz:2010gu} (for
other studies of such NP, see Refs.~\cite{Laplace:2002ik, Ligeti:2004ak, 
Agashe:2005hk, Bona:2005eu, Ligeti:2006pm, Isidori:2010kg, Ligeti:2010ia}).  We
use standard SM notation for the inputs, even for  quantities affected by NP in
$\Delta F=2$ whose measurements should be reinterpreted to include NP
contributions (e.g. $\alpha$, $\beta$, $\beta_s$). We consider 2003 (before the
first measurements of $\alpha$ and $\gamma$) and 2013 (as of the FPCP 2013
conference), and two future epochs, keeping in mind that any estimate of future
progress involves uncertainties on both experimental and theoretical sides.  Our
Stage~I projection refers to a time around or soon after the end of LHCb
Phase~I, corresponding to an anticipated 7\,fb$^{-1}$ LHCb data and 
5\,ab$^{-1}$ Belle~II data, towards the end of this decade.  The Stage~II
projection assumes 50\,fb$^{-1}$ LHCb and 50\,ab$^{-1}$ Belle~II data, and
probably corresponds to the middle of the 2020s, at the earliest.  Estimates of
future experimental uncertainties are taken from Refs.~\cite{Bediaga:2012py,
Aushev:2010bq, Meadows:2011bk, Bona:2007qt}.  (Note that we display the units as
given in the LHCb and Belle~II projections, even if it makes some comparisons
less straightforward; e.g., the uncertainties of both $\beta$ and $\beta_s$ will
be $\sim 0.2^\circ$ by Stage~II.)  For the entries in Table~\ref{bigtable} where
two uncertainties are given, the first one is statistical (treated as Gaussian)
and the second one is systematic (treated through the Rfit
model~\cite{ckmfitter}).   Considering the difficulty to ascertain the breakdown
between statistical and systematic uncertainties in lattice QCD inputs for the
future projections, for simplicity, we treat all such future uncertainties as
Gaussian.

The fits include the constraints from the measurements of $A_{\rm
SL}^{d,s}$~\cite{Laplace:2002ik, Lenz:2010gu}, but not their linear
combination~\cite{Abazov:2011yk}, nor from $\Delta\Gamma_s$, whose effects on
the future constraints on NP studied in this paper are small.  While
$\Delta\Gamma_s$ is in agreement with the CKM fit~\cite{Lenz:2010gu}, there are
tensions for $A_{\rm SL}$~\cite{Abazov:2011yk}. The large values of $h_{s}$
allowed until recently, corresponding to $(M_{12}^s)_{\rm NP} \sim -2
(M_{12}^s)_{\rm SM}$, are excluded by the LHCb measurement of the sign of
$\Delta\Gamma_s$~\cite{Aaij:2012eq}.  We do not consider $K$ mixing for the fits
shown in this Section, as it may receive NP contributions unrelated to $B_d$ and
$B_s$ mixings in the general case considered in this section. 

Figure~\ref{CKMfit} shows the evolution of the constraints on $(\rhobar,
\etabar)$ in the presence of NP in both $B_{d}$ and $B_s$ meson mixings, for
2003, 2013, Stage~I, and Stage~II.\footnote{Considering anticipated results from
only one experiment, plots similar to Fig.~\ref{CKMfit}, and with a different
parameterization, Fig.~\ref{Bdfit}, appear in Refs.~\cite{Aushev:2010bq,
Bona:2007qt}.}  The main constraints on $\bar\rho$ and $\bar\eta$ come from the
tree-level inputs $\gamma$ and $|V_{ub}|$, and also from the combination
$\gamma(\alpha)=\pi-\beta-\alpha$ which is not affected by NP in $\Delta
F=2$~\cite{Soares:1992xi}.  This constraint is more precise than $\gamma$ itself
until Stage~I, but of similar precision by Stage~II.  The $\gamma$ and
$\gamma(\alpha)$ measurement constraints are known modulo $\pi$, leading to a
sign ambiguity in the determination of $\bar\rho$ and $\bar\eta$.\footnote{In
2013, the combined constraint from the $\pi\pi$, $\pi\rho$ and $\rho\rho$ data
allows a second solution for $\alpha$ near 0, with a lower significance than the
SM solution in Table~\ref{bigtable}~\cite{Charles:2004jd}. This second solution
is shown as the negative-slope $\gamma(\alpha)$ wedge in Fig.~\ref{CKMfit}, and
is ruled out once combined with the $\gamma$ constraint. We assume that this
low-significance solution will disappear with more data by Stage~I.}  The
intersection of the $\gamma$, $\gamma(\alpha)$ and $|V_{ub}|$ constraints yields
two 95.5\%\,CL regions in Fig.~\ref{CKMfit} (yellow for positive $\bar\rho$ and
$\bar\eta$,  mauve for negative $\bar\rho$ and $\bar\eta$) symmetric with
respect to the origin.  This degeneracy is lifted by the addition of the other
experimental inputs, in particular $A_{\rm SL}^d$, leading to a single and small
95.5\%\,CL region (in yellow) for $\bar\rho$ and $\bar\eta$. (In 2013, the
degeneracy is only partially lifted: the $\bar\rho<0$, $\bar\eta<0$ solution is
excluded at 68.2\%\,CL, but it is allowed at 95.5\%\,CL.)

\begin{figure*}[tb]
\includegraphics[width=.48\textwidth,clip,bb=15 15 550 470]{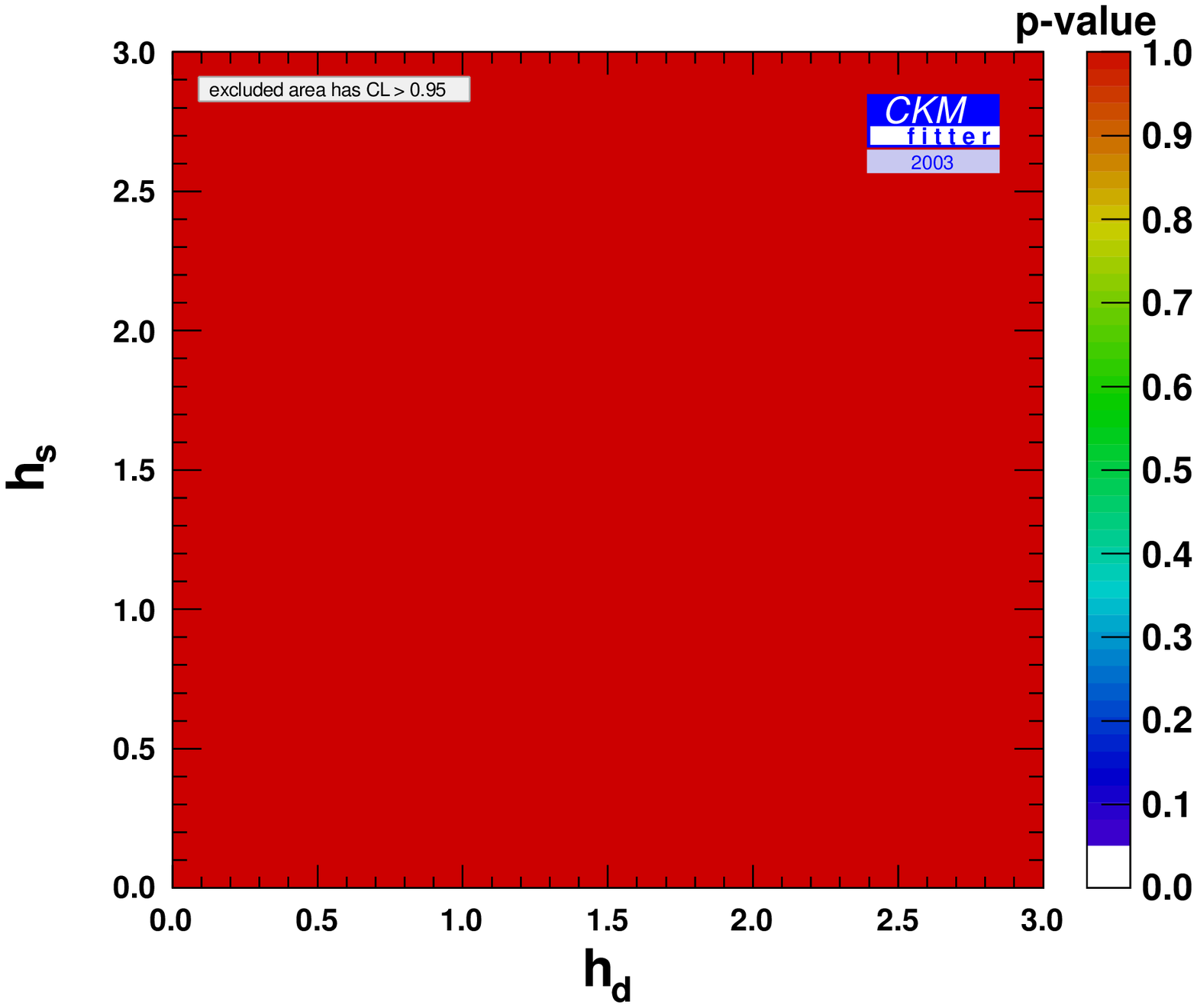}\hfill
\includegraphics[width=.48\textwidth,clip,bb=15 15 550 470]{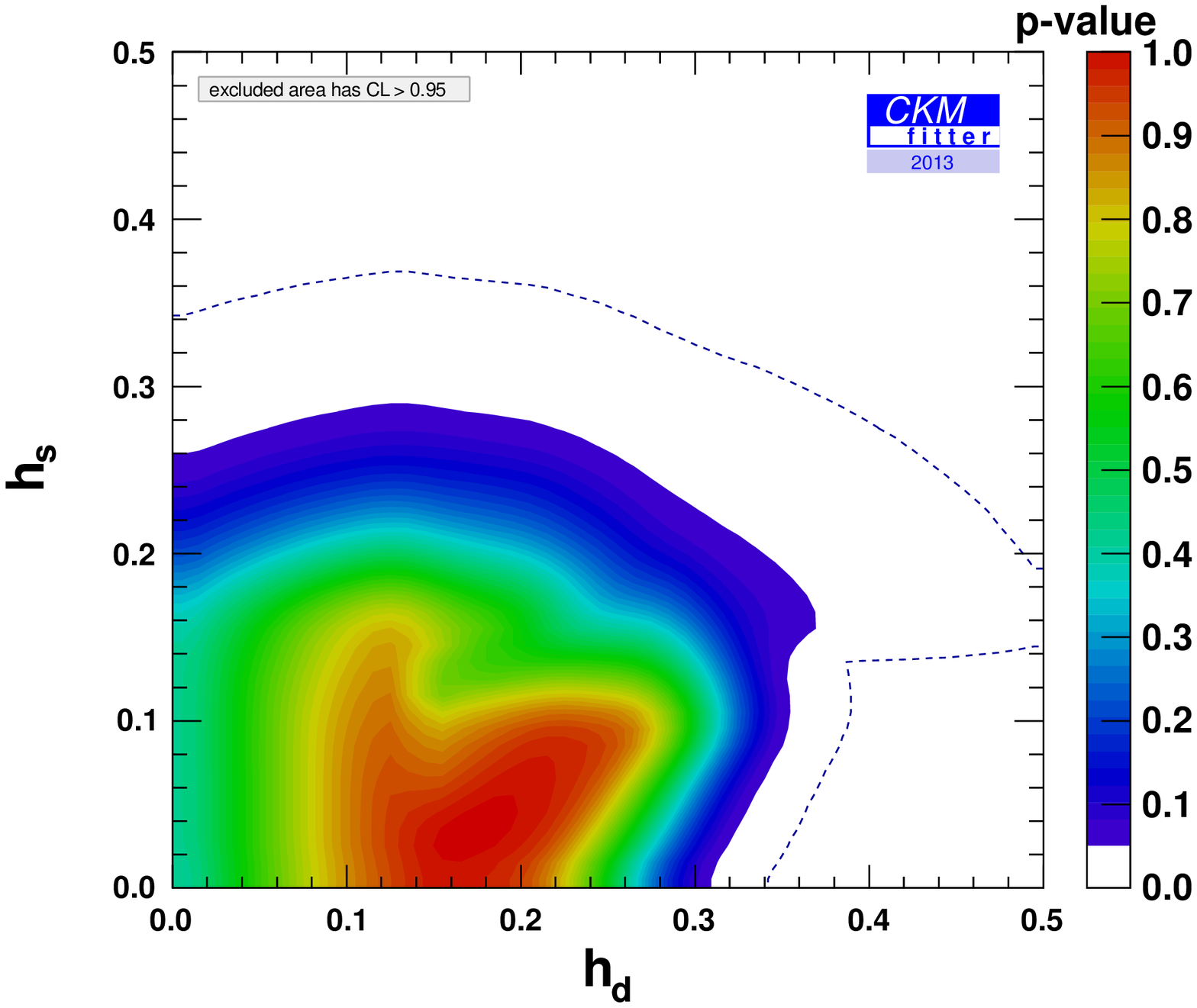}
\includegraphics[width=.48\textwidth,clip,bb=15 15 550 470]{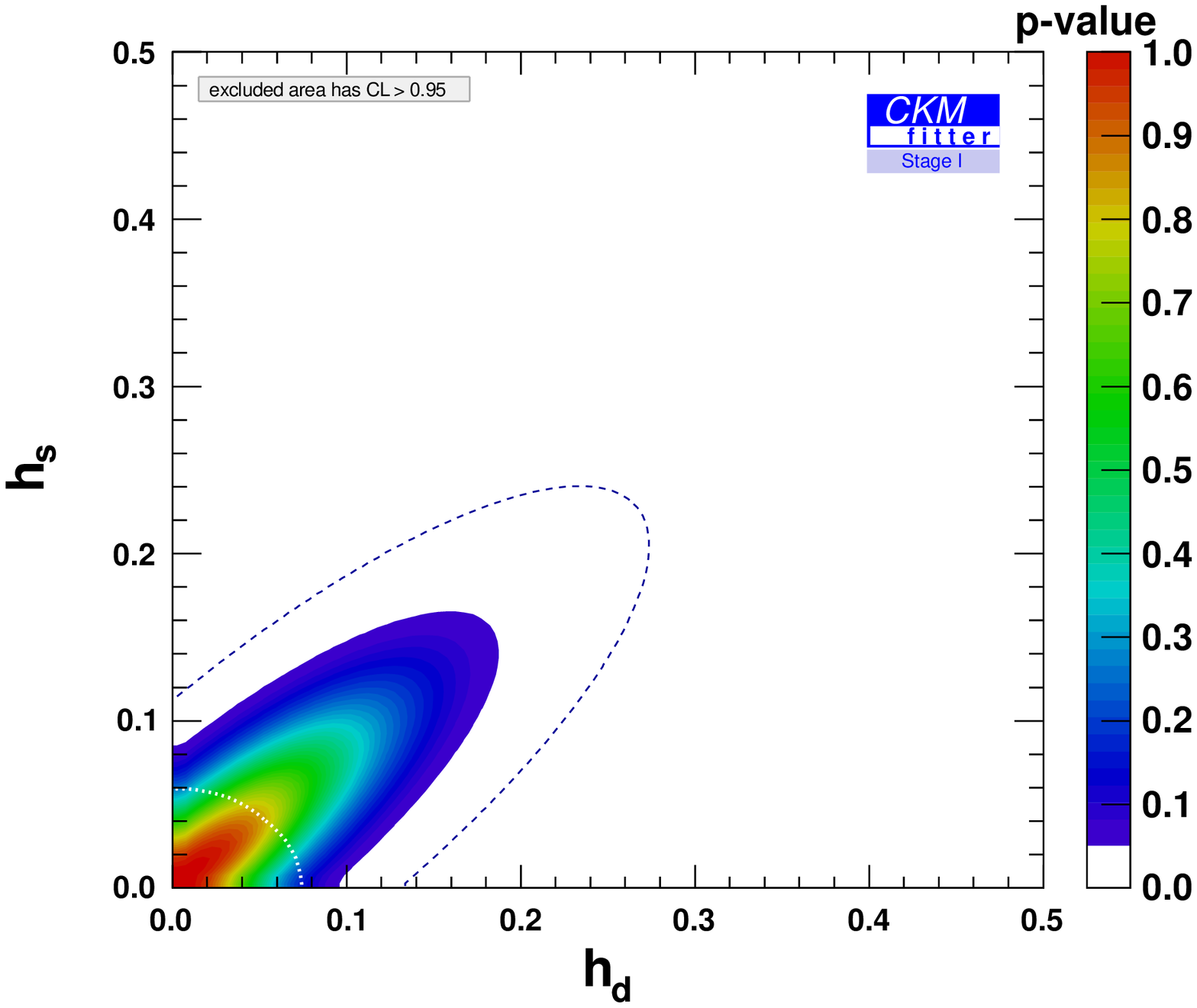}\hfill
\includegraphics[width=.48\textwidth,clip,bb=15 15 550 470]{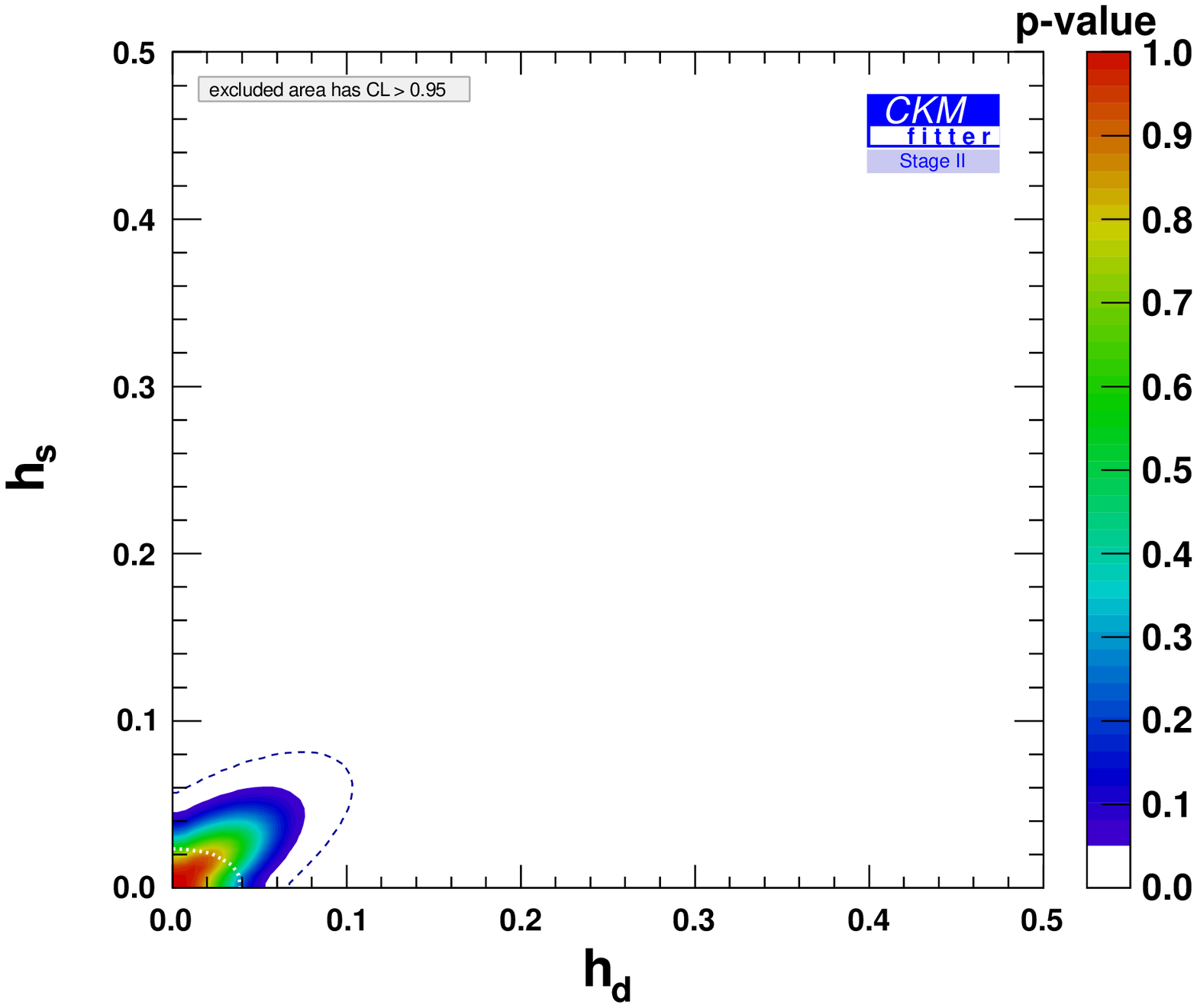}
\caption{The past (2003, top left) and present (top right) constraints on
$h_d-h_s$ in $B_{d}$ and $B_{s}$ mixings. The lower plots show future sensitivities for the
Stage~I and Stage~II scenarios described in the text, assuming measurements
consistent with the SM.  The dotted curves show the 99.7\%\,CL contours.  For 
Stage~I and Stage~II, the white dashed curves indicate the 95\%\,CL contours
obtained by setting theoretical uncertainties to zero.}
\label{hdhs}
\end{figure*}

Figures~\ref{Bdfit} and \ref{Bsfit} show the corresponding evolutions of the
constraints on $(h,\, \sigma)$ in the $B_d$ and $B_s$ meson systems.  Each plot 
is obtained by considering all the inputs in  Table~\ref{bigtable} and treating
$\rhobar$, $\etabar$, and the other physics parameters not shown as nuisance
parameters. This corresponds to the case of generic NP, ignoring possible
correlations between different $\Delta F=2$ transitions.   Since we are
interested in  the future sensitivity of LHCb and Belle~II to NP, for Stage~I
and Stage~II, we chose the central values of future measurements to coincide
with their SM predictions using the current best-fit values of $\rhobar$ and
$\etabar$. Thus, the future best fit corresponds to $h=0$. Figure~\ref{hdhs}
shows the projection on the $(h_d,\, h_s)$ plane.

Future lattice QCD uncertainties for Stage~I are taken from Refs.~\cite{USlqcd,
Ruth} (where they are given as expectations by 2018).  These predicted lattice
QCD improvements will be very important, mainly for the determination of
$|V_{ub}|$ and for the mixing matrix elements, $\langle B_q| (\bar b_L
\gamma_\mu q_L )^2 |\Bbar_q\rangle = (2/3)\, m_{B_q}^2 f_{B_q}^2 B_{B_q}$.  The
current expectation is that the uncertainties of $f_{B_q}$ will get below 1\%,
and may be significantly smaller than those of $B_{B_q}$.   The reduction of the
uncertainty of the latter to a similar level would be important.   Up to now,
due to the chiral extrapolations to light quark masses, more accurate results
were obtained for matrix elements involving the $B_s$ meson or for ratios
between $B_d$ and $B_s$ hadronic inputs, compared to the results for $B_d$
matrix elements.  This leads us to use the former quantities as our lattice
inputs for decay constants and bag parameters in Table~\ref{bigtable}. This
choice might not be the most suitable one in the future, due to improvements in
lattice results for light quarks. Concerning $|V_{ub}|$, it is reassuring that
2--3\% uncertainty should be obtainable from several measurements: $B\to
\tau\nu$, $B\to\mu\nu$, and $B\to\pi\ell\nu$ semileptonic decay.  For Stage~II,
we assumed some additional modest improvements in the lattice QCD inputs, which
are important mainly to constrain the MFV-like regions, $\sigma=0$ mod $\pi/2$. 
We studied the relative roles of the experimental measurements and the lattice
inputs at Stage~I and Stage~II. In Fig.~\ref{hdhs} the white dashed curves
indicate the 95\%\,CL contours obtained by setting the theoretical uncertainties
to zero, showing no correlation between $h_d$ and $h_s$. This is different from
a realistic situation (including theoretical uncertainties), in which case the
correlation between $h_{d}$ and $h_{s}$ in the Stage~I and II projections in
Fig.~\ref{hdhs} is driven by our current choice of ratios of $B_d$ and $B_s$
hadronic matrix elements as lattice inputs. This may not reflect the way lattice
results will improve in the future, and correlations will affect the shape of
the allowed regions in those plots.

From the discussion in the introduction, one may think that $\rhobar$ and
$\etabar$ are determined mostly by SM tree-level processes ($|V_{ub}/V_{cb}|$
and $\gamma$ from $B\to DK$ decays), while the additional loop-level observables
in the standard CKM fit constrain the NP.  In particular,  $\Delta M_{d,s}$,
$\sin 2\beta_{d,s}$, and $\alpha$ would constrain $h_{d,s}$ and $\sigma_{d,s}$,
while $\epsilon_{K}$ constrains $h_{K}$ and $\sigma_{K}$. This simple separation
of SM and NP has not been possible yet, given the large uncertainty of $\gamma$
compared to the combination, $\gamma(\alpha) \equiv \pi-\beta-\alpha$, which is
independent of NP in the classes of models under
consideration~\cite{Soares:1992xi}.  (Note that in the determination of $\alpha$
from $B\to \rho\rho,\, \rho\pi,\, \pi\pi$, an isospin analysis is used to remove
the penguin contribution.  To use this measurement to constrain new physics in
mixing, one has to assume that NP conserves isospin, which holds in most
scenarios, and is strongly supported by data.)  As one can clearly see from
Table~\ref{bigtable} and Fig.~\ref{CKMfit}, when the direct measurement of
$\gamma$ becomes as precise as $\pi-\beta-\alpha$ in the future, the separation
of the two sectors will be simpler to understand, even in a combined SM\,+\,NP
fit.

For our analysis, precise determination of CKM parameters from tree-level
measurements is essential, as illustrated in Fig.~\ref{CKMfit}. Depending on
future experimental results, the tension  between inclusive and exclusive
$|V_{ub}|$ (and $|V_{cb}|$) determinations might remain a cause for
concern~\cite{Beringer:1900zz}. The CKM part of our analysis also relies on the
expectation that the determination of $\gamma$ will indeed reach the $1^\circ$
level.  (For $\alpha$, a comparison of the $\rho\rho,\, \pi\pi,\, \rho\pi$
results may help to constrain the effects of isospin breaking and to reach the
expected accuracy.)

One can see from Figs.~\ref{Bdfit} and \ref{Bsfit} that recent LHCb measurements
have imposed comparable constraints on NP in $B_{s}$ mixing to those in the
$B_{d}$ system.  This qualitative picture will continue to hold for Stage~I and
Stage~II.  At Stage~I, we will have $h_{d,s}\lesssim 0.1$ for generic NP phases,
with an improvement by an additional factor of more than two at Stage~II. This
is not surprising, as the uncertainties on $\beta$ and $\beta_{s}$ will be
comparable, and improvements in the determination of $\rhobar$ and $\etabar$
from $\gamma$ and $|V_{ub}/V_{cb}|$ will affect the constraints on the two
systems in a similar way.  It is also interesting to see that the MFV regions
($\sigma_{d,s}=0$ mod $\pi/2$) will be less constrained also in the future.
Figure~\ref{hdhs} provides a different view of these results, by showing the
magnitudes of NP allowed in the $B_d$ vs.\ $B_s$ systems. 

\begin{figure*}[tb]
\includegraphics[width=.48\textwidth,clip,bb=15 15 550 470]{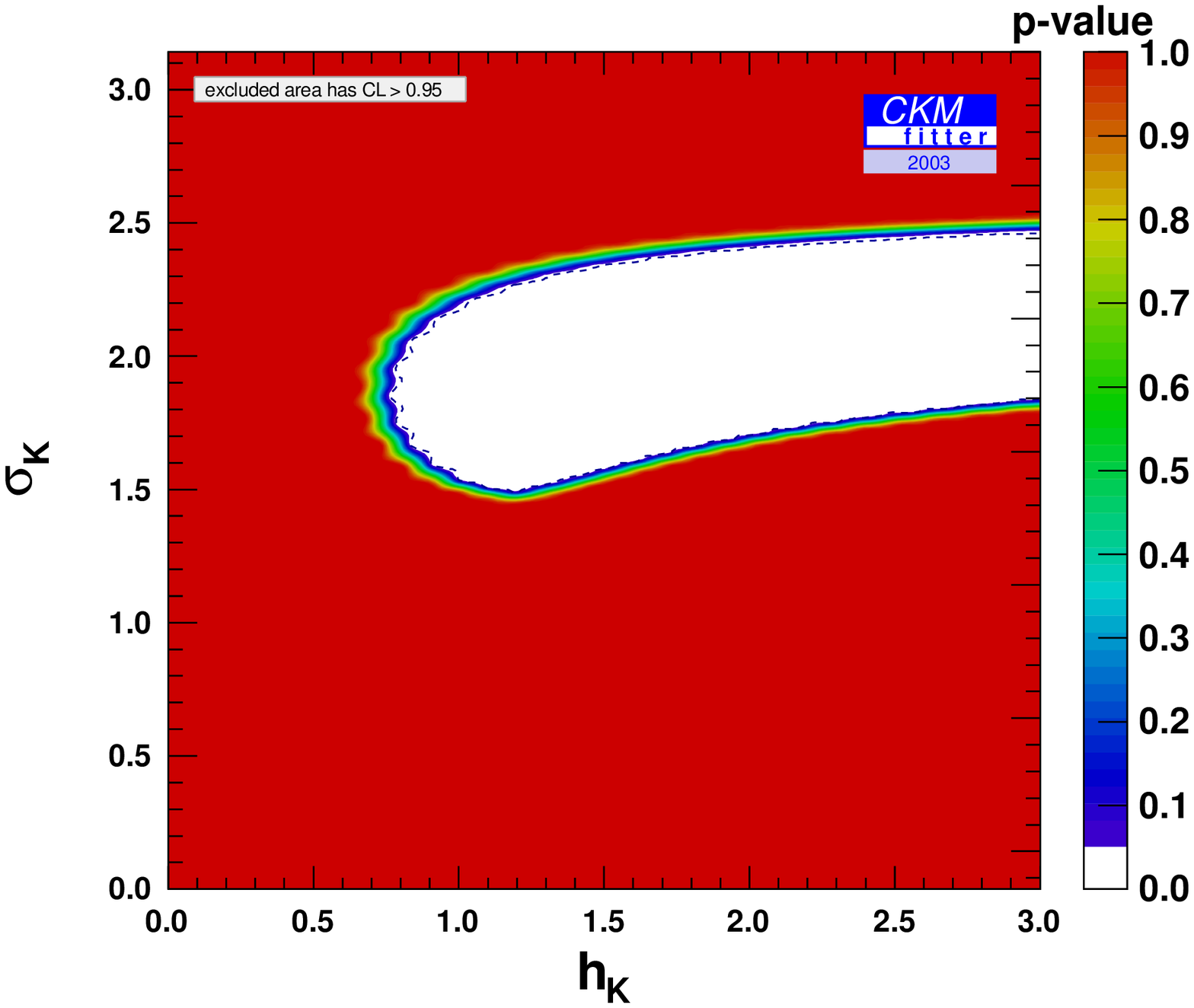}\hfill
\includegraphics[width=.48\textwidth,clip,bb=15 15 550 470]{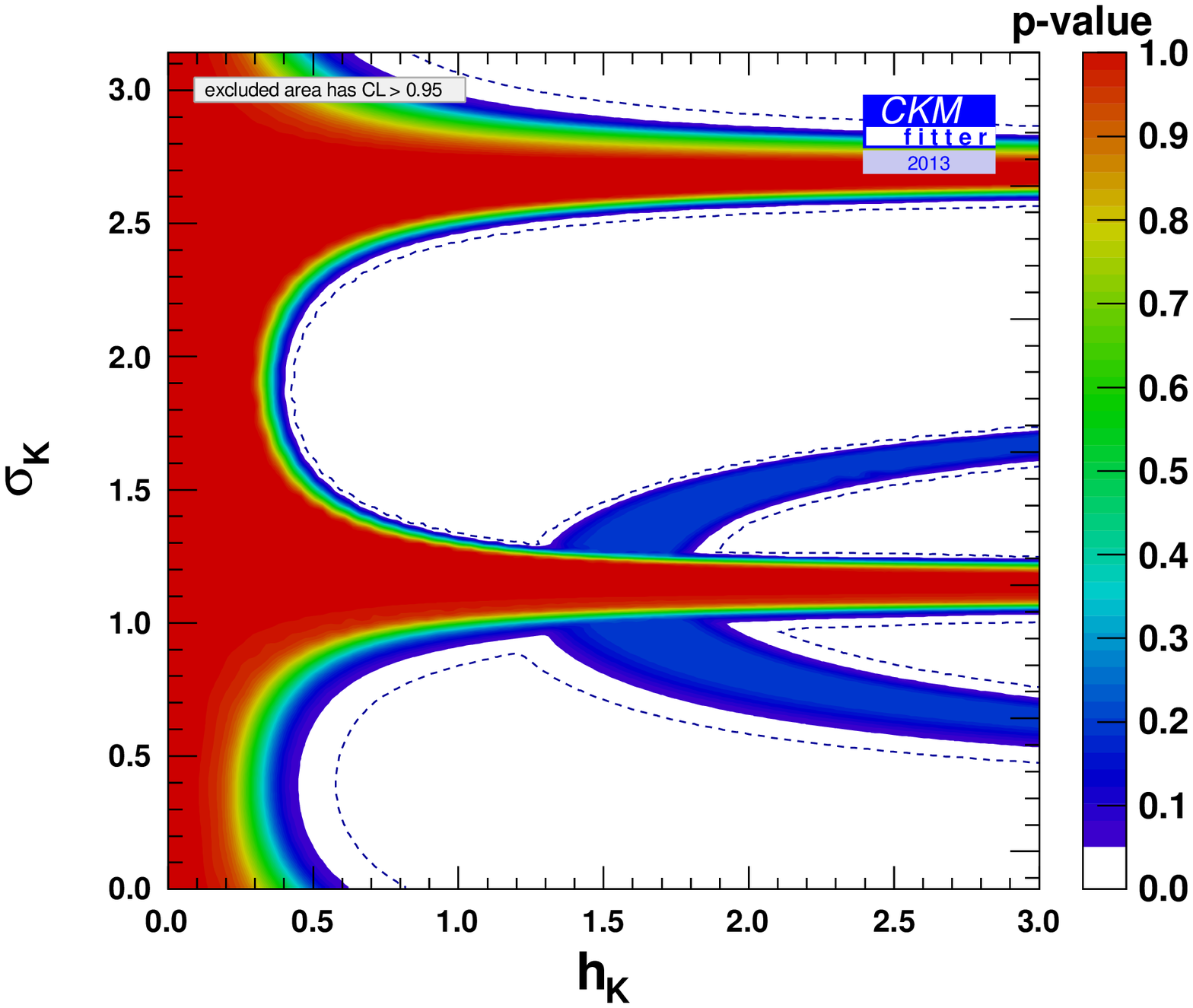}
\includegraphics[width=.48\textwidth,clip,bb=15 15 550 470]{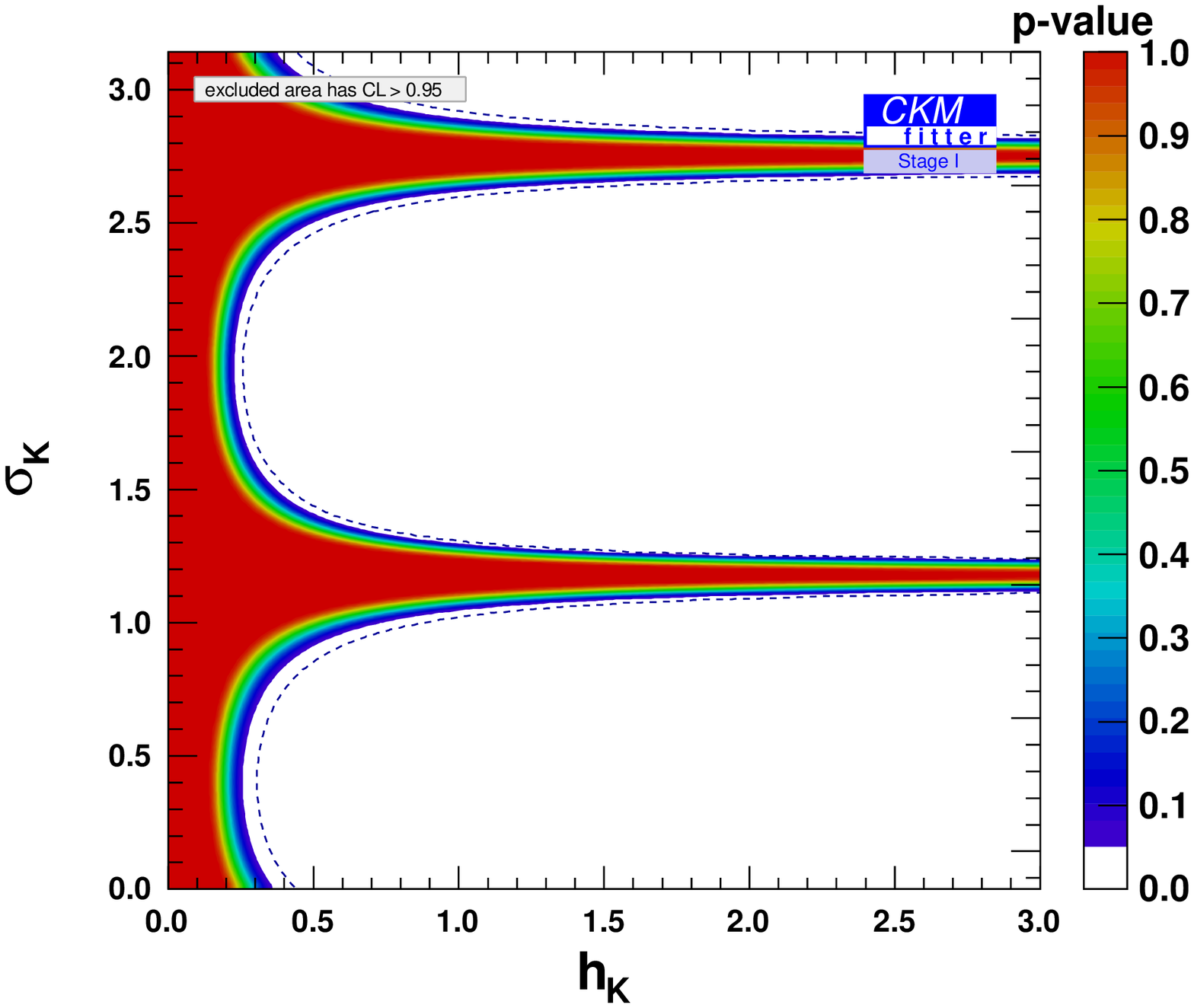}\hfill
\includegraphics[width=.48\textwidth,clip,bb=15 15 550 470]{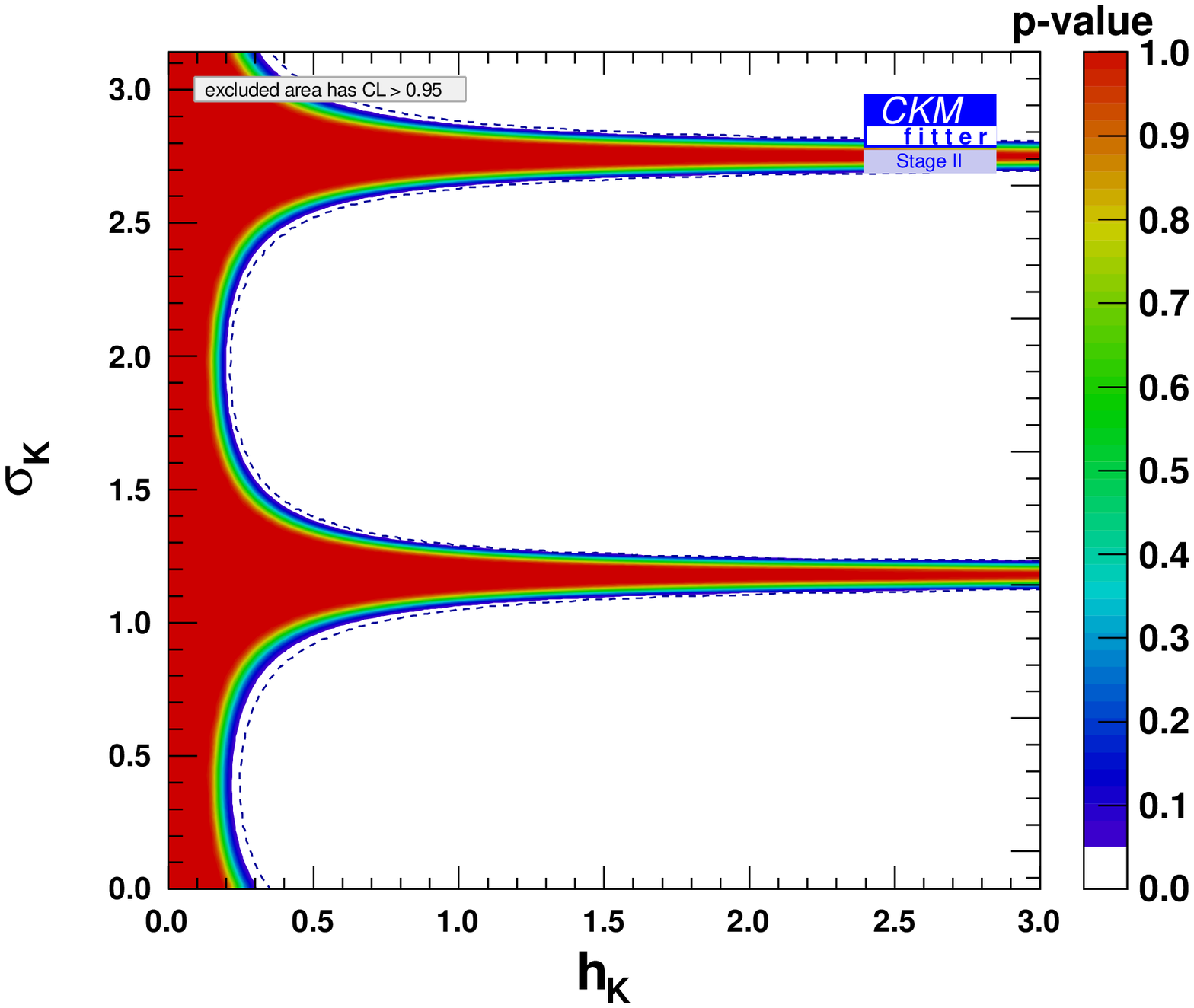}
\caption{The past (2003, top left) and present (top right) constraints on
$h_K-\sigma_K$ in $K$ mixing. The lower plots show future sensitivities for the
Stage~I and Stage~II scenarios described in the text, assuming measurements
consistent with the SM.  The dotted curves show the 99.7\%\,CL contours.}
\label{hKsK}
\end{figure*}

The better than factor-of-four improvement in the sensitivity to $h_{d,s}$ from
the current constraints to Stage~II, more than doubles the energy range probed
by these observables, and parallels the improvements in the high energy reach of
the LHC, going from LHC7 to LHC14.  If NP contains the same CKM suppressions of
$\Delta F=2$ transitions as those present for the SM contributions, typical for
models with nontrivial flavor structure in the LHC energy range, the scales
probed by the mixing constraints are
\beq\label{scalesCKM}
\Lambda \sim 17\,\TeV \ \ (\mbox{$B_d$})\,,
\qquad
\Lambda \sim 19\,\TeV \ \ (\mbox{$B_s$})\,.
\eeq
Here we used Eq.~(\ref{hnumeric}) with $|C_{ij}| = |\lambda^t_{ij}|$, and the
95\%\,CL bounds, $h_d < 0.07$ and $h_s < 0.06$ from Figs.~\ref{Bdfit}
and~\ref{Bsfit}.  If, instead, we use $|C_{ij}| = 1$ (corresponding to
non-hierarchical NP contributions), the probed scales are
\beq\label{scales1}
\Lambda \sim 2 \times 10^3\,\TeV \ \ (\mbox{$B_d$})\,,
\qquad 
\Lambda \sim 5 \times 10^2\,\TeV \ \ (\mbox{$B_s$})\,.
\eeq
Equation~(\ref{scalesCKM}) implies that LHCb and Belle~II will probe new
particles with CKM-like couplings with masses, $M$, in the 10--20\,TeV range if
they contribute at tree level (i.e., $\Lambda \sim M$), and in the 1--2\,TeV
range if they enter with a loop suppression (i.e., $\Lambda \sim 4\pi M$). 
Considering color factors, RGE effects, etc., which can differ for other
operators, one sees that these constraints are in the ballpark of gluino masses
explored at LHC14~\cite{ATLAS:2013hta}.

\section{\boldmath Including new physics in $K$ mixing}
\label{sec:npK}

Next, we consider the neutral kaon system, in addition to $B_d$ and $B_s$. We
only include the constraint from $\epsilon_{K}$, since there are large
uncertainties in the long-distance contribution to $\Delta m_{K}$ (for the same
reason, we do not study $D$-meson mixing). Figure~\ref{hKsK} shows the evolution
of the constraints on NP in $K$ mixing.  Larger values of $h_{K}$ for certain
values of the $CP$-violating phase will still be allowed, even at Stage~II. Due
to the presence of only one observable, $\epsilon_{K}$, constraining two
parameters, $h_K$ and $\sigma_K$, such ``throats'' cannot be eliminated. They
correspond to the values for which the imaginary part of the NP contribution
vanishes, that is $\sigma_{K}\sim \pi-\beta^{\rm SM}$ or $\pi/2 -\beta^{\rm
SM}$, where $\beta^{\rm SM}$ is the value of the true CKM $\beta$ angle shown in
Figure~\ref{CKMfit}.  In the 2013 plot, the two additional branches with low
p-values correspond to the less favored second solution for the CKM parameters
$\rhobar<0$, $\etabar<0$.

NP contributions as large as $30\%$ of the SM $tt$ contribution will be allowed
in the future, even in the MFV case, as can be seen by considering the
$\sigma_K=0,\pi/2$ values in the Stage~II plot in Fig.~\ref{hKsK}. Note that the
improvement from Stage~I to Stage~II is much less significant than the one from
the current status to Stage~I. Indeed, despite the almost factor-of-two
improvement on the uncertainty on $B_{K}$ and the improvements on $\rhobar$ and
$\etabar$, other parameters entering $\epsilon_K$ are not expected to have
similar improvements, as shown in Table~\ref{bigtable}. This includes the
uncertainty associated with higher-order terms in the OPE emphasized in
Ref.~\cite{Buras:2010pza}, and higher-order QCD corrections discussed in
Refs.~\cite{Brod:2010mj,Brod:2011ty} (in particular for the $cc$ contribution).

In many scenarios with TeV-scale NP, the constraints from kaon mixing provide
the strongest constraints to date, especially for the case of chirality-flipping
left-right ($LR$) operators, due to chiral enhancements in the matrix elements
and stronger QCD running. This situation will be maintained in the Stage~II era
as well, with comparable constraining power for non-$LR$ NP, and a significant
advantage of the kaon system over the $B_{d,s}$ systems in constraining
chirality-flipping operators. Furthermore, if NP is decoupled from the weak
scale and carries unsuppressed flavor violation (e.g., intermediate-scale split
supersymmetric scenarios~\cite{ArkaniHamed:2012gw}), the kaon system will
provide the most stringent probe (or the first place where a deviation can be
observed), since it carries the strongest CKM suppression in the SM.

\begin{figure}[tb]
\includegraphics[width=\columnwidth,clip,bb=20 15 550 390]{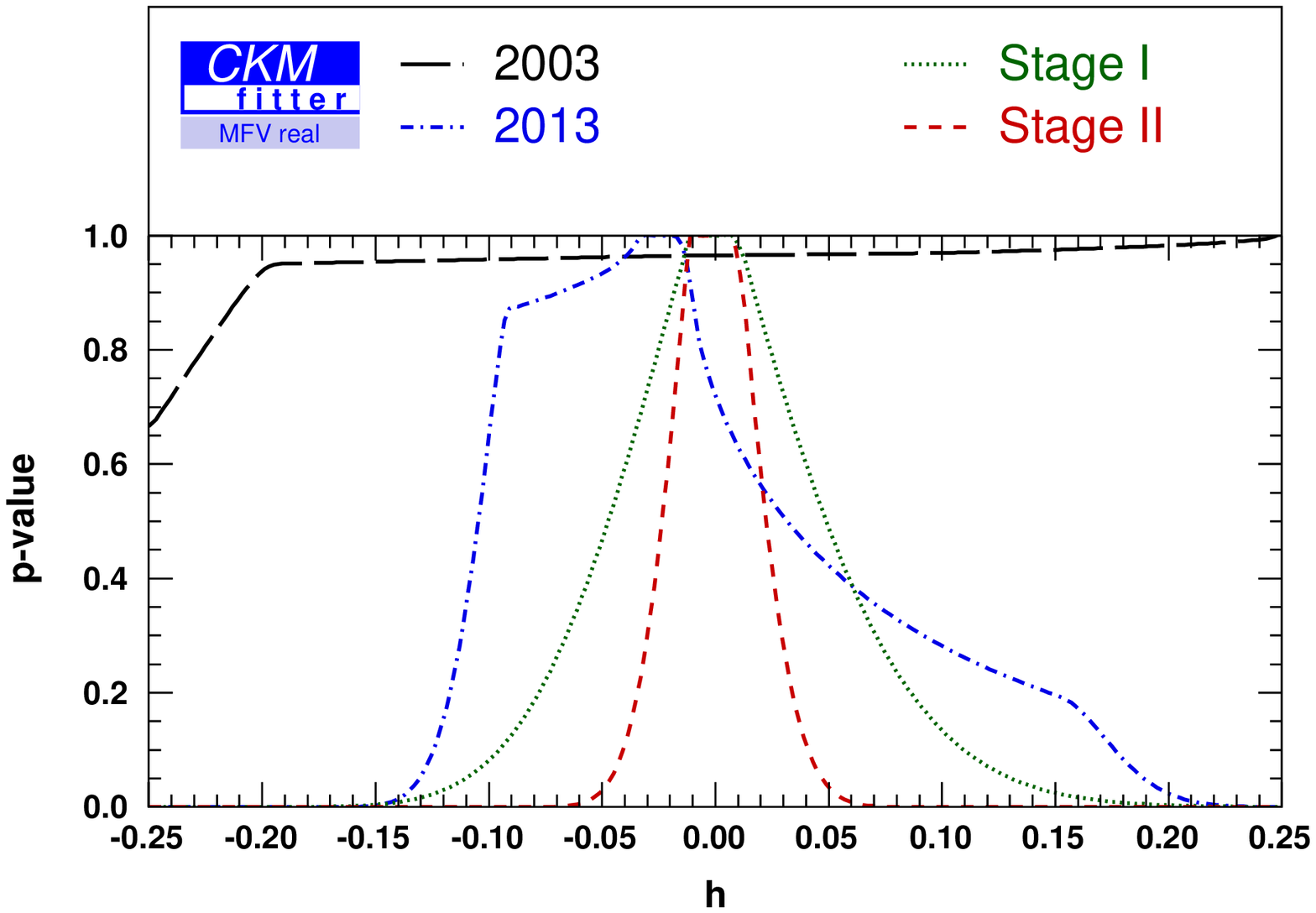}
\caption{Constraints on $h \equiv h_d\, e^{2i\sigma_d} = h_s\, e^{2i\sigma_s} =
h_K\, e^{2i\sigma_K}$ in MFV scenarios, in which $\sigma_d =\sigma_s =
\sigma_K=0$ (mod $\pi/2$), for the different epochs considered.}
\label{MFV}
\end{figure}

We next consider more specific NP scenarios, where the contributions to the
different neutral-meson systems are correlated. In the MFV case mentioned
in Sec.~\ref{sec:mixing}, 
\beq\label{MFVrel}
h_d = h_s = h_K\,, \qquad
0 = \sigma_d =\sigma_s = \sigma_K \ (\mbox{mod } \pi/2)\,.
\eeq
Figure~\ref{MFV} shows the $p$-values for the real (positive or negative)
$h \equiv h_d\, \exp(2i\sigma_d)$ in 2003, 2013, Stage~I, and Stage~II.

\begin{figure}[tb]
\includegraphics[width=\columnwidth,clip,bb=15 15 550 470]{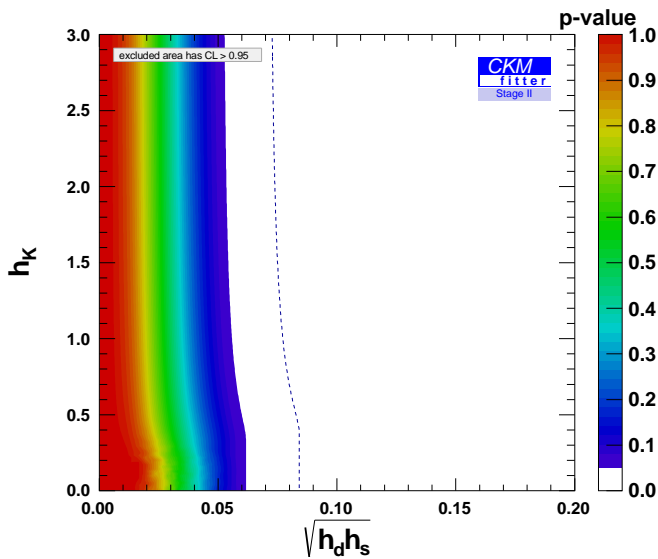}
\caption{Correlations between limits on NP in $K$, $B_d$ and $B_s$ mixing, at
``Stage~II'', in classes of models where flavor violation in $K$ mixing proceeds
dominantly via the third generation: $\sigma_{K}=\sigma_{d}-\sigma_{s}$, while
$h_{K,d,s}$ are kept free. The dotted curve shows the 99.7\%\,CL contour.}
\label{hkhds}
\end{figure}

Additional particularly interesting scenarios are those in which
the dominant effects are mediated by the third generation, motivated by the
natural stabilization of the electroweak scale, and those in which the
approximate horizontal $U(2)^3$ symmetry of the SM, induced by
$m_{u,c}/m_{t}\ll 1$ and $m_{d,s}/m_{b} \ll 1$, also applies to the NP
contributions~\cite{Barbieri:2011ci, Barbieri:2012uh}. 
In the first  case, the NP contribution to kaon mixing is attained via mixing
with the third generation, and is therefore related to those in $B_{d}$ and
$B_{s}$ mixings. In a fundamental theory representing this scenario the mixing
parameters $C_{ij}$ in the kaon sector, similar to Eq.~(\ref{operator}), will be
the product of those entering the $B_{d}$- and $B_{s}$-mixing expressions, up to
small corrections. Therefore, there is a correlation among the phases,
\beq\label{G3rel}
\sigma_K = \sigma_d - \sigma_s\,.
\eeq
On the other hand, the magnitudes of the NP contributions, $h_{K,d,s}$ also
depend on the typical mass scale, coupling constants, and kinematic function,
represented by $\Lambda$ in Eq.~(\ref{operator}). Thus, in general, 3rd
generation mediation in the kaon system does not imply a relation between
$h_{K}$ and $h_{d,s}$. The constraint on such models is shown in
Fig.~\ref{hkhds}, for the future Stage~II scenario. Mild correlations between
the limits on the magnitudes of NP in $B_{d,s}$ and $K$ mixings arise due to the
relations on the $CP$ phases $\sigma_{i}$ described above, and to a lesser
extent via $\rhobar$ and $\etabar$. The plot is easily understood: the largest
NP contribution in $B_{d,s}$ mixing is allowed for $\sigma_{d,s}\sim 0$ (mod
$\pi/2$), which is allowed for sufficiently small $h_{K}\lesssim 0.6$. The
presence of the ``throats" in Fig.~\ref{hKsK} allow larger values for $h_{K}$
for the non-$U(2)^3$ case, but at the price of not allowing $\sigma_{s,d} \sim
0$ (mod $\pi/2$), hence the (small) reduction in the allowed magnitude of NP in
the $B_{d,s}$ sector.  This effect can be more pronounced, if the actual future
data agrees less well with the SM, than assumed in this paper.

\begin{figure*}[tb]
\includegraphics[width=.48\textwidth,clip,bb=15 15 550 470]{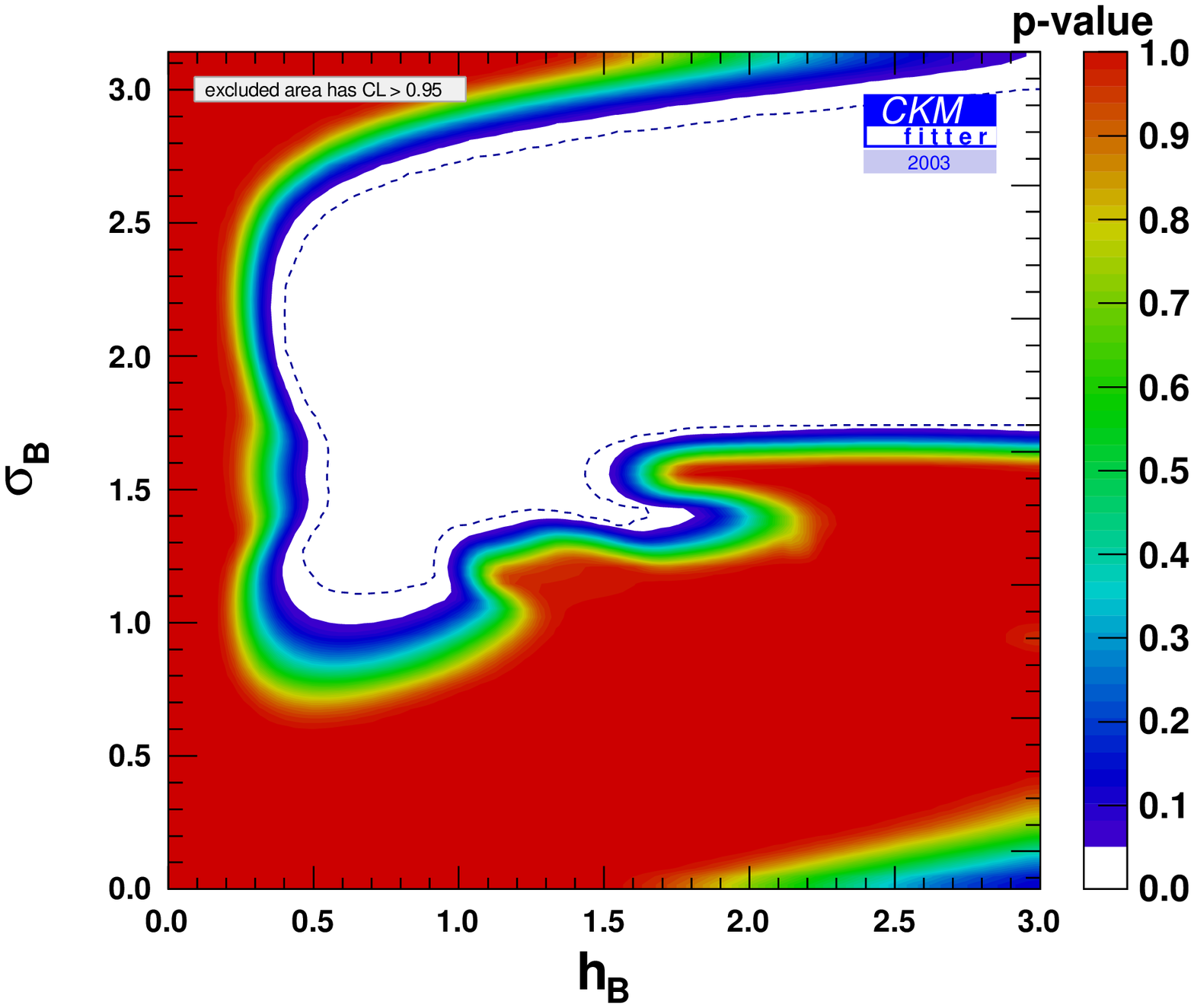}\hfill
\includegraphics[width=.48\textwidth,clip,bb=15 15 550 470]{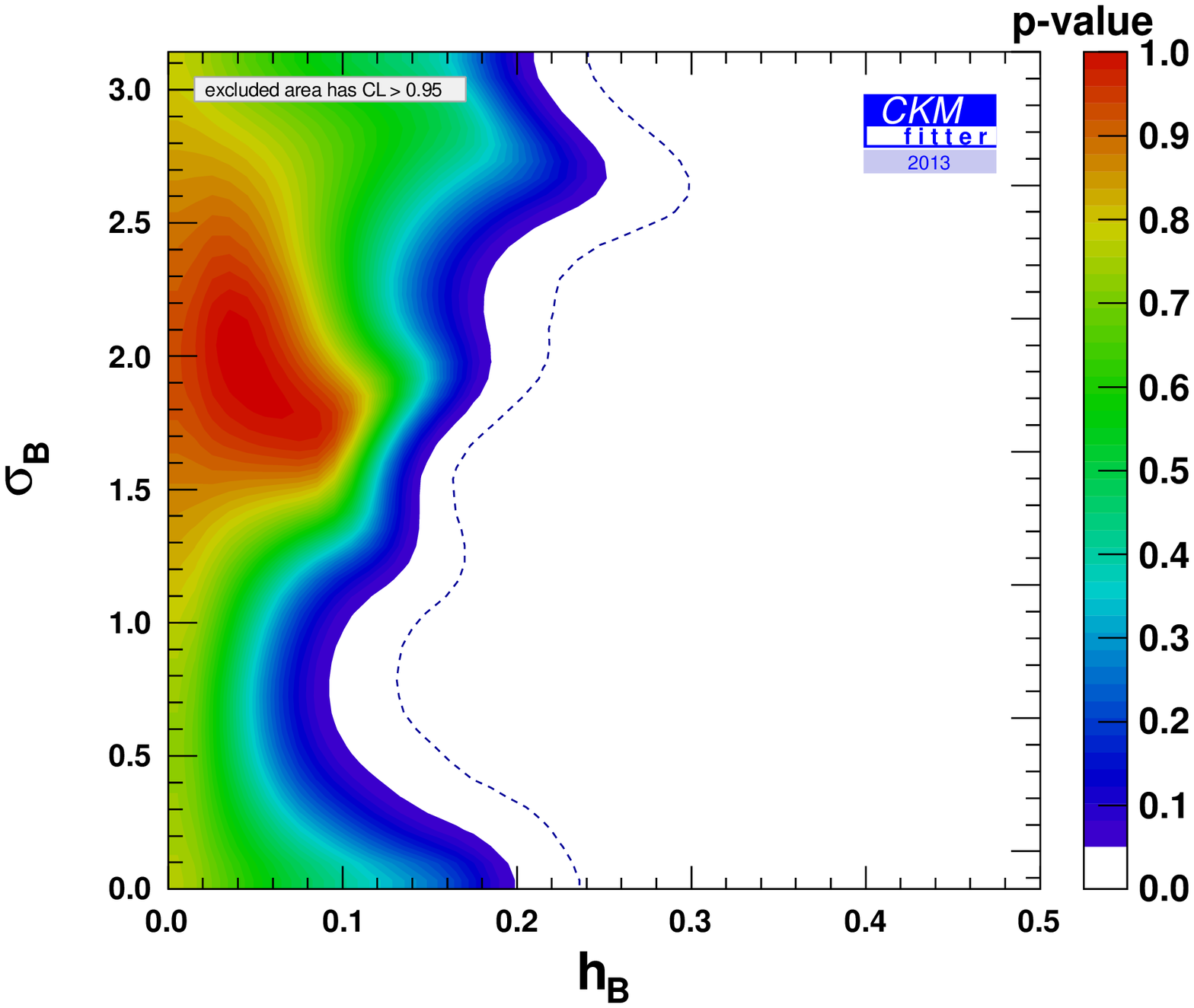}
\includegraphics[width=.48\textwidth,clip,bb=15 15 550 470]{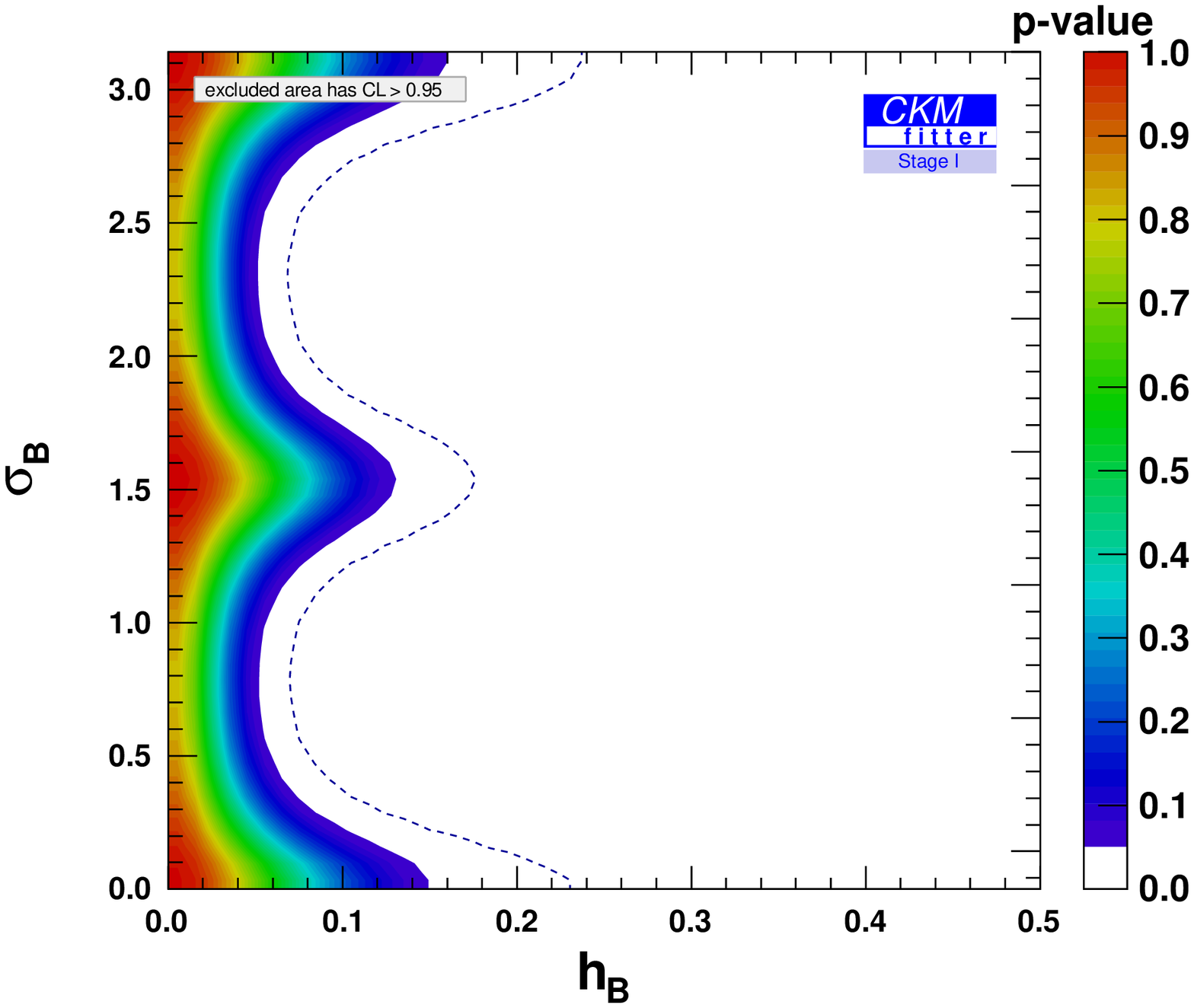}\hfill
\includegraphics[width=.48\textwidth,clip,bb=15 15 550 470]{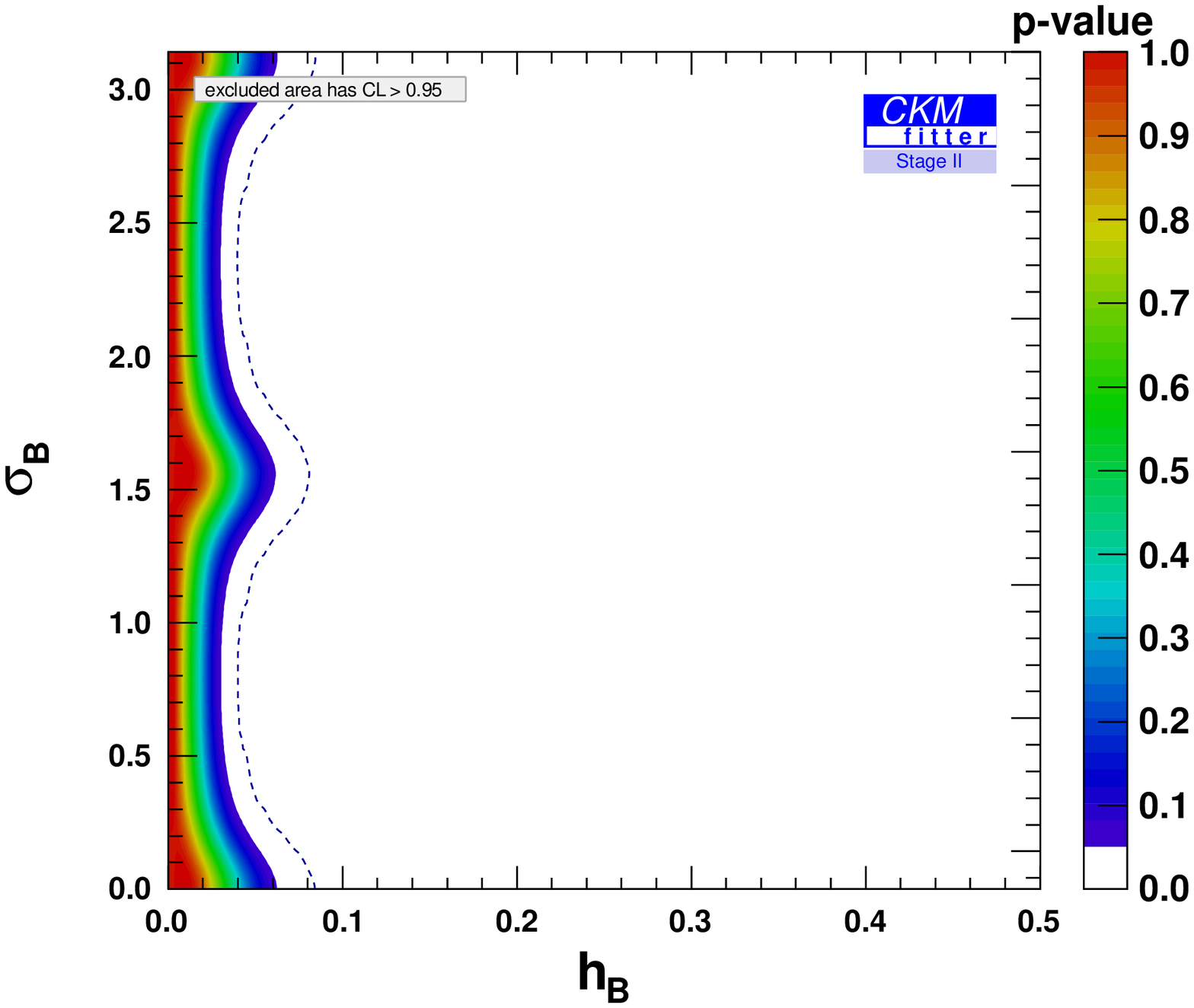}
\caption{The past (2003, top left) and present (top right) constraints on
$U(2)^3$ scenarios, where $h_B\equiv h_{d}=h_{s}$, $\sigma_B\equiv
\sigma_{d}=\sigma_{s}$. The lower plots show future sensitivities for the
Stage~I and Stage~II scenarios described in the text, assuming measurements
consistent with the SM.  The dotted curves show the 99.7\%\,CL contours.}
\label{hBsB}
\end{figure*}

\begin{figure*}[tb]
\includegraphics[width=\columnwidth,clip,bb=15 15 550 470]{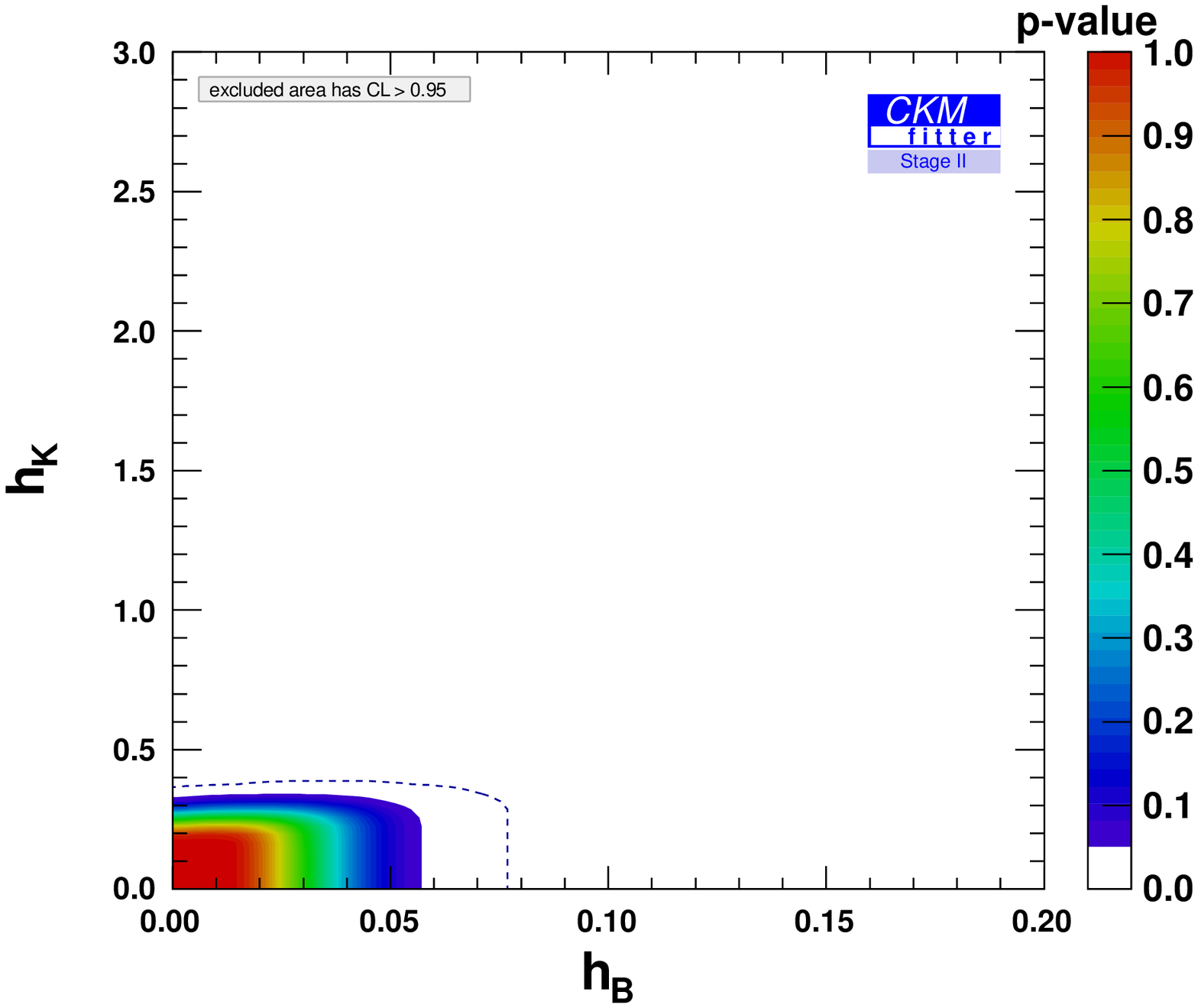}
\caption{Correlations between limits on NP in $K$ and $B_{d,s}$ mixing, at
``Stage~II'', in minimal $U(2)^3$ models. This fit corresponds to
$h_{d}=h_{s}\equiv h_{B}$, $\sigma_B \equiv \sigma_{d}=\sigma_{s}$, and
$\sigma_{K}=0$.  The dotted curve shows the 99.7\%\,CL contour.}
\label{hkhB}
\end{figure*}

In the case of the minimal $U(2)^3$ class of models~\cite{Barbieri:2012bh}, the
NP contributions to $B_{d}$ and $B_{s}$ should be equal. Furthermore, minimality
implies that the bulk of the NP contribution in the kaon sector is controlled by
the same spurions as in the $B_{d,s}$ sectors via 3rd generation mediation,
``23--31''. Therefore, one has
\beq\label{U2rel}
h_B \equiv h_d = h_s\,, \qquad \sigma_B \equiv \sigma_d = \sigma_s\,, 
\qquad \sigma_K =0\,.
\eeq
The constraints on such scenarios are shown in Figs.~\ref{hBsB} and \ref{hkhB}.
In Fig.~\ref{hBsB} the minimal $U(2)^3$ scenario is shown in the $h_B-\sigma_B$
plane.  While the 2003 and 2013 fits show interesting patterns arising from the
combination of the $B_{d,s}$ constraints, the future projections for the
$U(2)^3$ models look very similar to Figs.~\ref{Bdfit} and \ref{Bsfit}.

The correlation between the limits on the magnitudes of NP in $B_{d,s}$ and $K$
mixings in the minimal $U(2)^3$ case is shown in Fig.~\ref{hkhB} in the
$h_{K}$--$h_{B}$ plane.  Similar considerations as in Fig.~\ref{hkhds} apply
here.  As can be seen in Fig.~\ref{hKsK}, the constraint $\sigma_{K}=0$ limits
the size of $h_{K}\lesssim 0.25$.

In the case of generic $U(2)^3$ models, which allow additional NP
contributions in the kaon system unrelated to those in the $B_{d,s}$ systems,
$h_d=h_s$ and $\sigma_d = \sigma_s$ are maintained, but the correlation with the
$K$ systems is lost.  Therefore, the constraints in Fig.~\ref{hkhB} no longer
apply, while those in Fig.~\ref{hBsB} are still valid.

Constraints on NP in $K$ mixing will improve if lattice QCD gives a precise SM
calculation of $\Delta m_K$~\cite{{Christ:2012se}}.  For ${\rm Re}(M_{12}^K)$ in
the SM, the ratio of the $tt$ and $cc$ contributions is about 0.5\%, so a 1\%
calculation of $\Delta m_K$ could exclude $h_K \gtrsim 2$.  Lattice QCD progress
may also reduce the uncertainty in the higher order terms in $\epsilon_{K}$
discussed in Ref.~\cite{Buras:2010pza}, improving the overall constraints.  Due
to its unpredictability, we do not include possible improvements in this term
($\kappa_\epsilon$) in our Stage~I and II fits.  Even assuming a much reduced
uncertainty of $\eta_{cc}$, $\pm0.2$ instead of $\pm0.76$ at NNLO now (see
Ref.~\cite{Buras:2013raa}), would only improve the bounds on $h_K$ shown in
Fig.~\ref{hKsK} slightly; e.g., at Stage~II for $\sigma_K=0$, from $h_K < 0.31$
to $h_K < 0.24$.

In certain classes of models, improvement in sensitivity compared to
Fig.~\ref{hKsK} can also arise from future measurements of $K^+\to
\pi^{+}\nu\bar \nu$ and $K_L\to \pi^{0} \nu \bar \nu$~\cite{Grossman:1997sk}.
These decays are also sensitive to NP in $s \rightarrow d$ penguins, which can
be parameterized by another magnitude $h_{K}^{(\Delta S=1)}$ and phase
$\sigma_{K}^{(\Delta S=1)}$; thus the difference of the number of observables
vs.\ NP parameters will not change.  However, in certain well-motivated
scenarios, $\sigma_{K}^{(\Delta S=1)}=\sigma_{K}$~\cite{Agashe:2005hk}, or
$h_{K}^{(\Delta S=1)} \sim 0$, and in such cases including future data on these
rare decays will improve the sensitivity to NP.

\section{Summary and outlook}

We studied the anticipated future improvements in the constraints on NP in
$B_d$, $B_s$, and $K$ mixings.  We found that if no NP signal is seen, the
bounds on $h_d$ and $h_s$ will improve by about a factor of 5.  This corresponds
to probing NP at scales more than a factor of two higher than currently (for a
fixed set of couplings).  Interestingly, compared to the allowed regions to
date, we expect the MFV-like regions, $\sigma=0$ (mod $\pi/2$), to be nearly as
strongly constrained as those with generic NP phase in the future.  Our results
for the future sensitivity to a NP contribution given by Eq.~(\ref{operator}) in
$B_d$ and $B_s$ mixings at Stage~II are summarized in Table~\ref{scalestable}.
For $K$ mixing, the large $h_K$ regions in Fig.~\ref{hKsK} complicate the
interpretation in terms of NP scales.  If we assume that lattice QCD will
exclude $h_K>2$ as discussed in Sec.~\ref{sec:npK}, we get sensitivity up to 3
TeV (0.3 TeV) at tree level (one loop) for CKM-like couplings, and up to
$9\times 10^3$ TeV ($7\times 10^2$ TeV) at tree level (one loop) for
non-hierarchical couplings.

\begin{table}[b!]
\begin{tabular}{c|c|c|c}
\hline\hline
\multirow{2}{*}{Couplings}  &  NP loop  &  
  \multicolumn{2}{c}{Scales (in TeV) probed by}\\
&  order  &  $B_d$ mixing  &  $B_s$ mixing   \\
\hline\hline
$|C_{ij}| = |V_{ti}V_{tj}^*|$ & tree level &  17  & 19   \\ 
\cline{2-4}
(CKM-like)  &  one loop  &  1.4  &  1.5   \\
\hline
$|C_{ij}| = 1$  &  tree level  &  $2\times 10^3$  &  $5\times 10^2$ 
    \\
\cline{2-4}
(no hierarchy)  &  one loop &  $2\times 10^2$  &  40 \\
\hline\hline
\end{tabular}
\caption{The scale of the operator in Eq.~(\ref{operator}) probed by $B_d$ and
$B_s$ mixings at Stage~II (if the NP contributions to them are unrelated). The
impact of CKM-like hierarchy of couplings and/or loop suppression is indicated.}
\label{scalestable}
\end{table}

\begin{figure*}[tb]
\includegraphics[width=.48\textwidth,clip,bb=15 15 550 470]{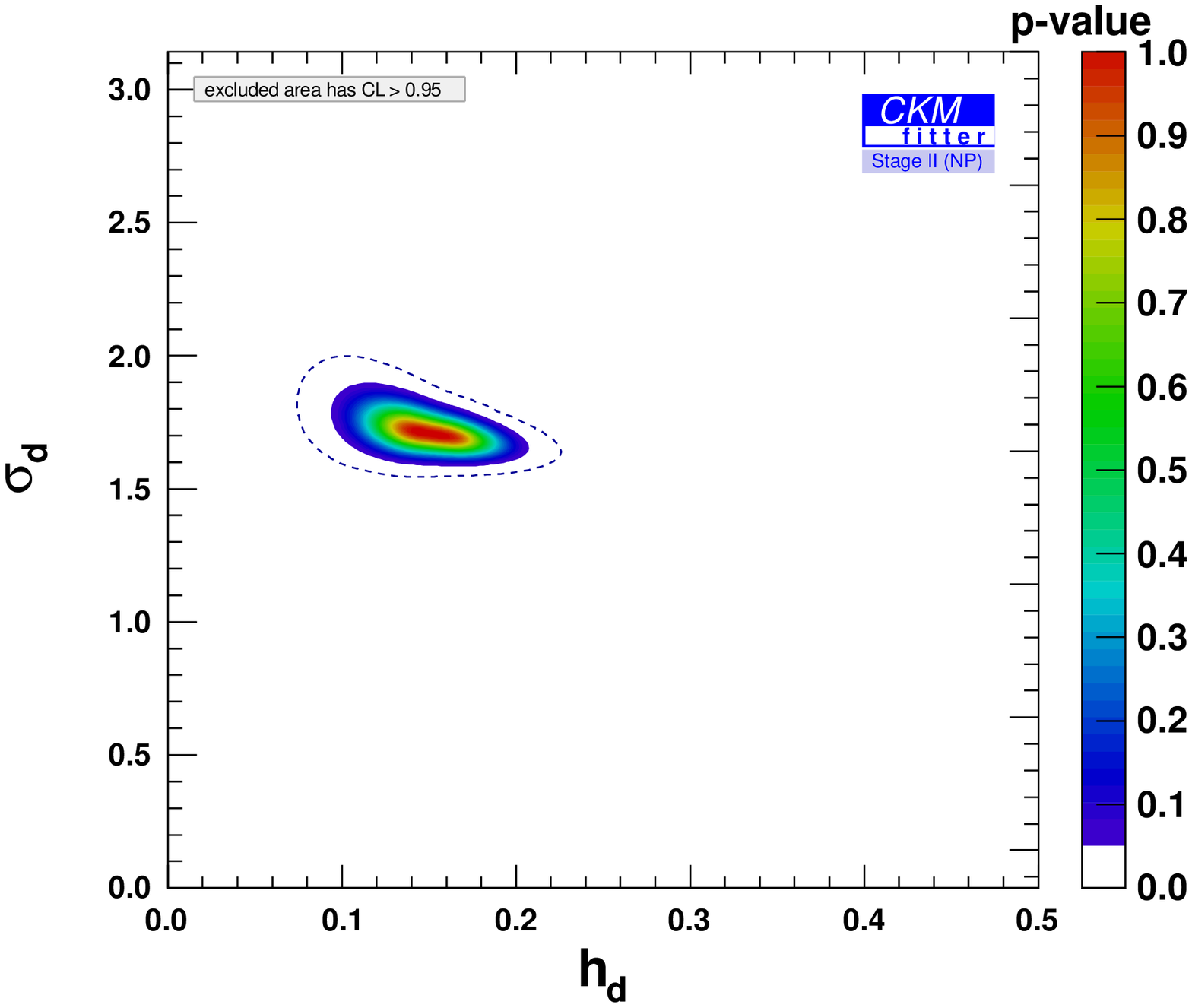}\hfill
\includegraphics[width=.48\textwidth,clip,bb=15 15 550 470]{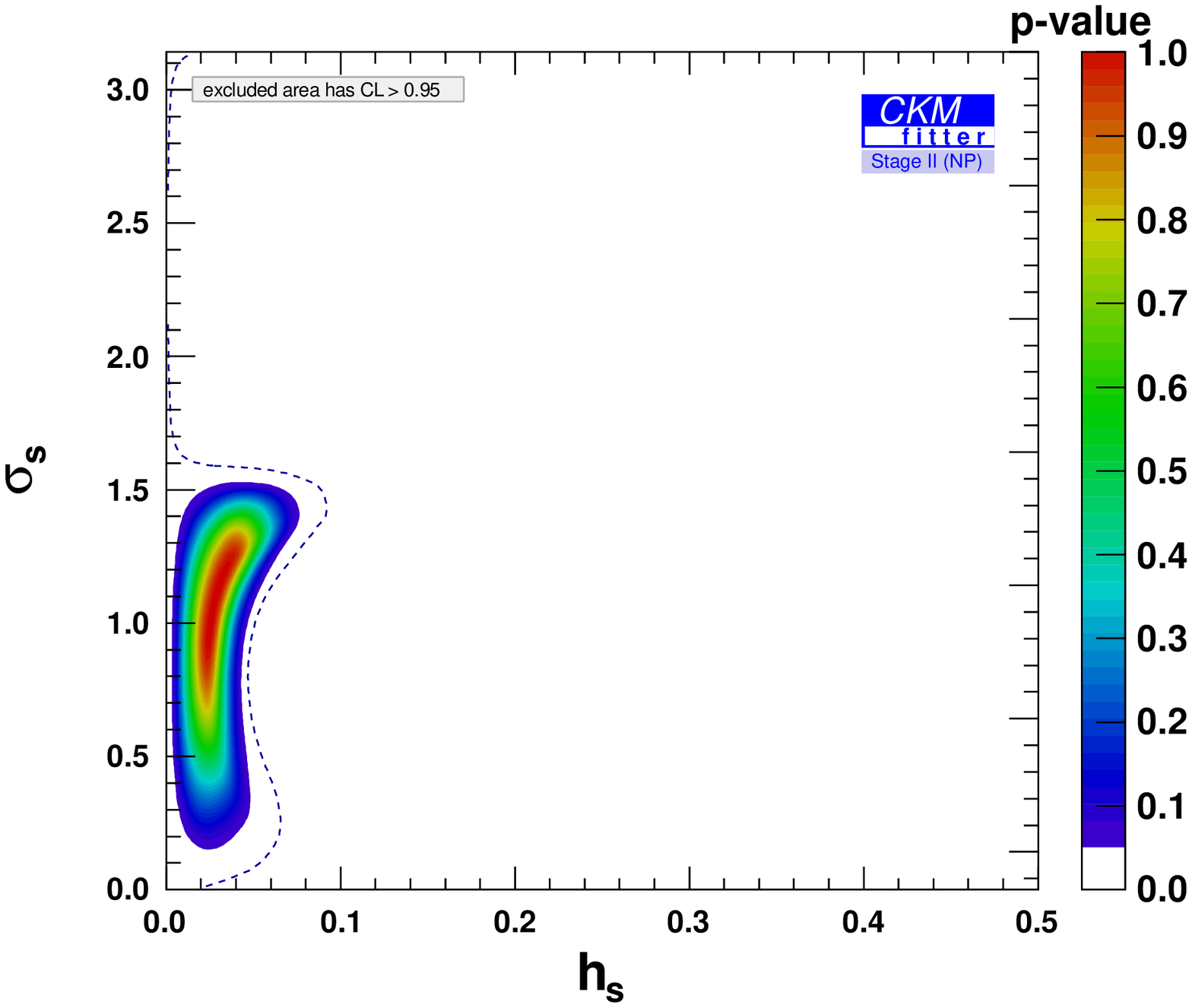}
\caption{Hypothetical Stage~II fits for NP, assuming that all future
experimental results correspond to the current best-fit values of $\rhobar$,
$\etabar$, $h_{d,s}$ and $\sigma_{d,s}$ (with measurement uncertainties as
given in Table~\ref{bigtable}, but different central values).}
\label{NPsignal}
\end{figure*}

So far in this paper we assumed that future measurements agree with the SM
predictions. However, future data can not only set better bounds on NP, they may
also reveal deviations from the SM.  This is illustrated in Fig.~\ref{NPsignal},
where we set $\rhobar$, $\etabar$, $h_{d,s}$ and $\sigma_{d,s}$ to their current
best-fit values (allowing for NP in $\Delta F=2$), and performed a fit assuming
for all future measurements the corresponding central values, but uncertainties
as given in Table~\ref{bigtable} for Stage~II.  While any assumption about
possible future NP signals includes a high degree of arbitrariness,
Fig.~\ref{NPsignal} may give an impression of the sensitivity to reveal a
deviation from the SM.

Similar predictions could be made for many other higher dimension
flavor-changing operators.  The $\Delta F=1$ observables dominated by one-loop
contributions in the SM probe different NP contributions.  Such analyses have
been performed for $b\to s\gamma$, $b\to s \ell^{+}\ell^{-}$, etc.~\cite{DF=1}. 
The progress for the constraints imposed by some of these observables,
especially those corresponding to not yet observed processes, will be greater
than those for $B_d$ and $B_s$ mixings studied in this paper.  This example is
particularly interesting, as many NP models do predict an effect which may be
observable in the coming decade.  Furthermore, $\Delta F=2$ generically provides
the strongest constraints for high-scale models with unsuppressed flavor
violation, while still providing competitive constraints for lower scale NP
(where flavor transitions are parametrically suppressed as in the SM). Finally,
the significant improvements on the bounds in the $h-\sigma$ planes for $B_d$
and $B_s$ mixings give an impressive yet conservative illustration of the
anticipated future progress coming from the LHCb upgrade and the Belle~II
experiment.

\acknowledgments

We thank Riccardo Barbieri, Filippo Sala, and St\'ephane T'Jampens for helpful
comments.
ZL and MP were supported in part by the Office of Science, Office of High Energy
Physics, of the U.S.\ Department of Energy under contract DE-AC02-05CH11231.
We would like to thank all members of the CKMfitter group for suggestions
on various aspects of this article.


\begin{thebibliography}{99}

\bibitem{Kobayashi:1973fv}
  M.~Kobayashi and T.~Maskawa,
  Prog.\ Theor.\ Phys.\  {\bf 49} (1973) 652.
  
\bibitem{Soares:1992xi}
  J.~M.~Soares and L.~Wolfenstein,
  Phys.\ Rev.\ D {\bf 47} (1993) 1021; 
  T.~Goto, N.~Kitazawa, Y.~Okada and M.~Tanaka,
  Phys.\ Rev.\ D {\bf 53} (1996) 6662
  [hep-ph/9506311]; 
  J.~P.~Silva and L.~Wolfenstein,
  Phys.\ Rev.\ D {\bf 55} (1997) 5331
  [hep-ph/9610208]; 
Y.~Grossman, Y.~Nir and M.~P.~Worah,
Phys.\ Lett.\ B {\bf 407} (1997) 307
[hep-ph/9704287].

\bibitem{Hocker:2006xb}
  A.~Hocker and Z.~Ligeti,
  Ann.\ Rev.\ Nucl.\ Part.\ Sci.\  {\bf 56} (2006) 501
  [hep-ph/0605217].

\bibitem{Charles:2004jd}
  J.~Charles {\it et al.}  [CKMfitter Group],
  Eur.\ Phys.\ J.\ C {\bf 41} (2005) 1
  [hep-ph/0406184];
and updates at \url{http://ckmfitter.in2p3.fr/}.

\bibitem{Ligeti:2004ak}
  Z.~Ligeti,
  Int.\ J.\ Mod.\ Phys.\ A {\bf 20} (2005) 5105
  [hep-ph/0408267].

\bibitem{Buras:2001ra} 
  A.~J.~Buras, S.~Jager and J.~Urban,
  Nucl.\ Phys.\ B {\bf 605}, 600 (2001)
  [hep-ph/0102316].

\bibitem{Fox:2007in} 
  P.~J.~Fox, Z.~Ligeti, M.~Papucci, G.~Perez and M.~D.~Schwartz,
  Phys.\ Rev.\ D {\bf 78}, 054008 (2008)
  [arXiv:0704.1482 [hep-ph]].

\bibitem{ckmfitter}
A.~Hocker, H.~Lacker, S.~Laplace and F.~Le Diberder,
Eur.\ Phys.\ J.\ C {\bf 21} (2001) 225
[hep-ph/0104062].

\bibitem{Charles:2011va}
  J.~Charles
  {\it et al.},
  Phys.\ Rev.\ D {\bf 84} (2011) 033005
  [arXiv:1106.4041 [hep-ph]].

\bibitem{Lenz:2010gu} 
  A.~Lenz {\it et al.},
  Phys.\ Rev.\ D {\bf 83}, 036004 (2011)
  [arXiv:1008.1593 [hep-ph]];
  A.~Lenz {\it et al.},
  Phys.\ Rev.\ D {\bf 86}, 033008 (2012)
  [arXiv:1203.0238 [hep-ph]].

\bibitem{Laplace:2002ik}
S.~Laplace, Z.~Ligeti, Y.~Nir and G.~Perez,
Phys.\ Rev.\ D {\bf 65} (2002) 094040
[hep-ph/0202010].

\bibitem{Agashe:2005hk}
  K.~Agashe, M.~Papucci, G.~Perez and D.~Pirjol,
  hep-ph/0509117.

\bibitem{Bona:2005eu} 
  M.~Bona {\it et al.}  [UTfit Collaboration],
  JHEP {\bf 0603}, 080 (2006)
  [hep-ph/0509219];
  JHEP {\bf 0803}, 049 (2008)
  [arXiv:0707.0636 [hep-ph]].

\bibitem{Ligeti:2006pm}
  Z.~Ligeti, M.~Papucci and G.~Perez,
  Phys.\ Rev.\ Lett.\  {\bf 97} (2006) 101801
  [hep-ph/0604112].

\bibitem{Isidori:2010kg}
  G.~Isidori, Y.~Nir and G.~Perez,
  Ann.\ Rev.\ Nucl.\ Part.\ Sci.\  {\bf 60} (2010) 355
  [arXiv:1002.0900 [hep-ph]].

\bibitem{Ligeti:2010ia}
  Z.~Ligeti, M.~Papucci, G.~Perez and J.~Zupan,
  Phys.\ Rev.\ Lett.\  {\bf 105} (2010) 131601
  [arXiv:1006.0432 [hep-ph]].

\bibitem{Aushev:2010bq}
  T.~Aushev
  {\it et al.},
  arXiv:1002.5012 [hep-ex].

\bibitem{Bediaga:2012py}
  R.~Aaij {\it et al.}  [LHCb Collaboration],
  Eur.\ Phys.\ J.\ C {\bf 73} (2013) 2373
  [arXiv:1208.3355 [hep-ex]].

\bibitem{USlqcd}
T.~Blum {\it et al.}, ``Lattice QCD at the Intensity Frontier"
\url{http://www.usqcd.org/documents/13flavor.pdf}.

\bibitem{Ruth}
We thank R.\ Van De Water for helpful correspondance about future lattice QCD
expectations.

\bibitem{Meadows:2011bk}
  B.~Meadows
  {\it et al.},
  arXiv:1109.5028 [hep-ex].

\bibitem{Bona:2007qt}
  M.~Bona {\it et al.}, 
  arXiv:0709.0451 [hep-ex].

\bibitem{Abazov:2011yk} 
  V.~M.~Abazov {\it et al.}  [D0 Collaboration],
  Phys.\ Rev.\ D {\bf 84}, 052007 (2011)
  [arXiv:1106.6308 [hep-ex]].

\bibitem{Aaij:2012eq} 
  R.~Aaij {\it et al.}  [LHCb Collaboration],
  Phys.\ Rev.\ Lett.\  {\bf 108}, 241801 (2012)
  [arXiv:1202.4717 [hep-ex]].

\bibitem{Beringer:1900zz}
See, e.,g., R.~Kowalewski and T.~Mannel, ``Determination of $V_{cb}$ and
$V_{ub}$"; in
J.~Beringer {\it et al.}  [Particle Data Group Collaboration],
  Phys.\ Rev.\ D {\bf 86} (2012) 010001.

\bibitem{ATLAS:2013hta} 
  ATLAS Collaboration,
  arXiv:1307.7292 [hep-ex].

\bibitem{Buras:2010pza} 
  A.~J.~Buras, D.~Guadagnoli and G.~Isidori,
  Phys.\ Lett.\ B {\bf 688}, 309 (2010)
  [arXiv:1002.3612 [hep-ph]].

\bibitem{Brod:2010mj}
  J.~Brod and M.~Gorbahn,
  Phys.\ Rev.\ D {\bf 82} (2010) 094026
  [arXiv:1007.0684 [hep-ph]].
  
\bibitem{Brod:2011ty}
  J.~Brod and M.~Gorbahn,
  Phys.\ Rev.\ Lett.\  {\bf 108} (2012) 121801
  [arXiv:1108.2036 [hep-ph]].
    
\bibitem{ArkaniHamed:2012gw} 
  N.~Arkani-Hamed, A.~Gupta, D.~E.~Kaplan, N.~Weiner and T.~Zorawski,
  arXiv:1212.6971 [hep-ph].

\bibitem{Barbieri:2011ci}
  R.~Barbieri, G.~Isidori, J.~Jones-Perez, P.~Lodone and D.~M.~Straub,
  Eur.\ Phys.\ J.\ C {\bf 71} (2011) 1725
  [arXiv:1105.2296 [hep-ph]].

\bibitem{Barbieri:2012uh} 
  R.~Barbieri, D.~Buttazzo, F.~Sala and D.~M.~Straub,
  JHEP {\bf 1207}, 181 (2012)
  [arXiv:1203.4218 [hep-ph]].

\bibitem{Barbieri:2012bh} 
  R.~Barbieri, D.~Buttazzo, F.~Sala and D.~M.~Straub,
  JHEP {\bf 1210}, 040 (2012)
  [arXiv:1206.1327 [hep-ph]].

\bibitem{Christ:2012se} 
  N.~H.~Christ, T.~Izubuchi, C.~T.~Sachrajda, A.~Soni and J.~Yu,
  arXiv:1212.5931 [hep-lat];
see also 
S.~Sharpe, talk at the ANL Intensity Frontier Workshop, 25--27 April 2013,
\url{https://indico.fnal.gov/getFile.py/access?contribId=126&sessionId=0&resId=0&materialId=slides&confId=6248},
and J.~Yu, talk at Lattice 2013, 29 July -- 3 August 2013.
\url{http://www.lattice2013.uni-mainz.de/presentations/7C/Yu.pdf}

\bibitem{Buras:2013raa} 
  A.~J.~Buras and J.~Girrbach,
  arXiv:1304.6835 [hep-ph].

\bibitem{Grossman:1997sk}
  Y.~Grossman and Y.~Nir,
  Phys.\ Lett.\ B {\bf 398}, 163 (1997)
  [hep-ph/9701313].

\bibitem{DF=1}
  T.~Hurth {\it et al.},
  Nucl.\ Phys.\ B {\bf 808} (2009) 326
  [arXiv:0807.5039 [hep-ph]];
%
  F.~Beaujean {\it et al.},
  JHEP {\bf 1208} (2012) 030
  [arXiv:1205.1838 [hep-ph]];
%
  S.~Descotes-Genon {\it et al.},
  JHEP {\bf 1106} (2011) 099
  [arXiv:1104.3342 [hep-ph]];
  JHEP {\bf 1301} (2013) 048
  [arXiv:1207.2753 [hep-ph]];
%
    S.~Descotes-Genon, J.~Matias and J.~Virto,
  arXiv:1307.5683 [hep-ph];
%
  W.~Altmannshofer and D.~M.~Straub,
  JHEP {\bf 1208} (2012) 121
  [arXiv:1206.0273 [hep-ph]];
  arXiv:1308.1501 [hep-ph].

%


\end{thebibliography}
\end{document}